\documentclass[fleqn,usenatbib]{mnras}
\usepackage{newtxtext,newtxmath}
\usepackage[T1]{fontenc}
\DeclareRobustCommand{\VAN}[3]{#2}
\let\VANthebibliography\thebibliography
\def\thebibliography{\DeclareRobustCommand{\VAN}[3]{##3}\VANthebibliography}

\usepackage{graphicx}	
\usepackage{amsmath}	
\usepackage{color}

\title{Halo Growth and Merger Rates as a Cosmological Test}
\author[Y. Amoura et al.]{Yuba Amoura,$^{1,2}$\thanks{E-mail: ayuba@uwaterloo.ca} 
Nicole E.~Drakos,$^{3}$
Anael Berrouet$^{2}$, 
James E.~Taylor$^{1,2}$\thanks{E-mail: taylor@uwaterloo.ca}
\\
$^{1}$Waterloo Centre for Astrophysics, University of Waterloo, Waterloo, Ontario N2L 3G1 Canada\\
$^{2}$Department of Physics and Astronomy, University of Waterloo, 200 University Avenue West, Waterloo, Ontario N2L 3G1, Canada\\
$^{3}$Department of Physics and Astronomy, University of Hawaii, Hilo, 200 W Kawili St, Hilo, HI 96720, USA
}

\date{December 2022}

\begin{document}
\label{firstpage}
\pagerange{\pageref{firstpage}--\pageref{lastpage}}
\maketitle
\begin{abstract}
Dark matter haloes grow at a rate that depends on the value of the cosmological parameters $\sigma_8$ and $\Omega_{\rm m}$ through the initial power spectrum and the linear growth factor. While halo abundance is routinely used to constrain these parameters, through cluster abundance studies, the halo growth rate is not. In recent work, we proposed constraining the cosmological parameters using observational estimates of the overall dynamical ``age'' of clusters, expressed, for instance, by their half-mass assembly redshift $z_{50}$. Here we explore the prospects for using the instantaneous growth rate, as estimated from the halo merger rate, from the average growth rate over the last dynamical time, or from the fraction of systems with recent episodes of major growth. We show that the merger rate is mainly sensitive to the amplitude of fluctuations $\sigma_8$, while the rates of recent growth provide constraints in the $\Omega_{\rm m}$--$\sigma_8$ plane that are almost orthogonal to those provided by abundance studies. Data collected for forthcoming cluster abundance studies, or studies of the galaxy merger rate in current and future galaxy surveys, may thus provide additional constraints on the cosmological parameters complementary to those already derived from halo abundance. 
\end{abstract}



\section{Introduction}

The standard Lambda Cold Dark Matter ($\Lambda$ CDM) cosmological model provides an extremely effective framework for understanding and predicting cosmological observations. As the accuracy of observational constraints increases, however, several small discrepancies have begun to challenge the success of the $\Lambda $CDM model. In particular, in measurements of the amplitude of density perturbations, (commonly represented by the parameter $\sigma_8$, the rms of density fluctuations smoothed on scales of 8\,Mpc/h), and the growth of perturbations (determined by the matter density parameter $\Omega_m$), a tension has emerged between results based on the CMB power spectrum \citep{Planck.2020, Aiola_ACT2020} and studies probing the late-time matter density field directly through weak gravitational lensing \citep[e.g.][]{Heymans2020}
or cluster abundance \citep[e.g.][]{Bocquet2019_CC, Abdullah2020}. 

This discrepancy, referred to as the ``$S_8$ tension'' in reference to the growth of structure parameter $S_8 \propto \sqrt{\Omega_m}\sigma_8$, has grown significantly in recent years as more precise weak-lensing studies have been released, in particular the Dark Energy Survey (DES Y3) \citep{Abbott2022_DESY3} and the Kilo Degree Survey (KiDS-1000) \citep{Heymans2021_KiDS1003x2pt}, and is about $\sim 3\sigma$ at the time of writing. Various solutions to the tension have been proposed, including systematic problems in the analysis \citep[e.g.][]{Sanchez2020}, biased cluster mass estimates \citep[e.g.][]{Douspis.2019, Debackere.2021} or modifications to the standard cosmological model \citep[e.g.][]{DiValentino2015, Bohringer2016, Planck2016_modifiedG, Heimersheim2020}, but it is not clear that any of these fully resolve the problem.
Given the persistent discrepancy, it is worth exploring other independent methods for estimating $\sigma_8$ and $\Omega_m$, to investigate all possible origins for the tension. 

In the standard cosmological model, dark matter becomes  non-relativistic (cold) at early times, and structures grow hierarchically after the initial gravitational collapse of peaks in the density field \citep{Joeveer&Einasto1978, Peebles.1980, White1991, Padmanabhan1993, Dodelson2003}.
Within this framework, the largest structures, galaxy clusters, are the last objects to assemble most of their mass, yet they also form around and thus probe the highest peaks in the initial density field. Consequently, present-day cluster abundance can be used to estimate the early abundance and subsequent growth of the density peaks, which in turn have a clear dependence on $\sigma_8$ and $\Omega_m$ \citep{Press.Schechter}. The cluster count method has been used extensively for several decades to constrain these parameters  \citep{Evrard1989, Henry&Arnaud1991, Lilje1992, Wang_1998, Abdullah2020}.

This cosmological test uses remarkably little information about individual clusters, requiring only their observed redshift and mass or mass proxy, as determined from observations in the X-ray \citep[e.g.][]{Henry.2009, Mantz.2010, Bohringer.2014}, weak-lensing surveys \citep[e.g.][]{Kacprzak2016}, optical galaxy surveys, or sub-mm imaging via the Sunyaev-Zel'dovich effect \citep[e.g.][]{deHaan.2016, Planck.2020}. Despite the wealth of information present in these observational data, structural features of clusters such as their concentration, substructure and shape have not been exploited for cosmological purposes. This is partly because of the difficulty of measuring, modelling and understanding these features, but also because the large data sets necessary to perform cosmological analyses using highly variable, complex properties such as cluster structure were previously unavailable. This situation is now changing rapidly, as forthcoming missions and surveys, including \emph{Euclid}, DESI \citep{DESI2016}, the \emph{Vera C. Rubin Observatory} \citep{LSST2009}, \emph{eRosita} \citep{Pillepich2012}, the \emph{Nancy Grace Roman Telescope}, or UNIONS \citep{UNIONS2020}, are expected to provide data for very large samples of galaxy clusters. Furthermore, new approaches to the analysis of complex nonlinear data, such as those associated with machine learning, are becoming more common. These new data sets and new analysis tools make cosmological analyses with cluster structural properties a promising avenue to explore.

The idea of using the cosmological dependence of cluster formation histories to constrain cosmology is not new, but was discussed in the literature three decades ago \citep{Richstone1992, Evrard1993, Mohr1995}. These original tests leveraged the fact that the structural properties of galaxy clusters are related to how relaxed they are -- their projected shape and non-axisymmetry -- and to the state of the Universe when they accreted their mass -- concentration -- \citep[see][for a review]{Taylor.2011}. Subsequent work has examined structural properties, showing that they are generally consistent with expectations from LCDM \citep[e.g.][]{Oguri2010, Sereno2018}, but has not used them to constrain cosmological parameters specifically. 

In \citet{Amoura2021} (paper I hereafter), we showed that for values of $\sigma_8$ varying between 0.75--0.85, the resulting median age of galaxy clusters, as expressed by the epoch $z_{50}$ by which a system had accreted half its final mass, would vary by more than 10\%. Combining accurate, unbiased measurements of structural parameters such as concentration for a large enough sample, such a difference could easily be distinguished in future cluster samples. More importantly, for clusters of mass $\sim10^{14}M_\odot/h$ at low redshift, the constraints obtained this way are orthogonal to the typical `banana-shaped' constraints following contours of constant $S_8$. 

While our previous work focussed on the overall `age' of clusters, i.e. some average measure such as $z_{50}$ defined over their whole accretion history, the instantaneous growth rate may sometimes be easier to determine from observations. This rate is reflected in halo merger rates, in the mean increase in mass over some recent interval of time, or in the fraction of systems that have recently experienced a large increase in mass.

Tests of halo growth or the merger rate could in principle be applied on any mass scale traced by visible matter. The merger rate on galaxy scales has been studied extensively both observationally and in simulations, using various tracers of merger activity, including close pairs of galaxies, starbursts, and morphologically distorted galaxies \citep[e.g.][]{Lotz.2011, Xu.2012, Mundy.2017}. Since the machinery for estimating halo merger rates is the same independent of scale, we will also consider galaxy-scale growth and merger rates, although constraining these with observations involves several additional challenges, as discussed in Section 4.

The outline of the paper is as follows. In Section 2, we use analytical models based on the Extended Press-Schechter (EPS) formalism to estimate how various measures of halo growth vary with $\Omega_m$ and $\sigma_8$. In Section 3 we compare these predictions to dark-matter-only $N$-Body simulations, and discuss the discrepancies between the two. In Section 4 we consider the prospects for measuring halo growth observationally, either directly on cluster scales, or indirectly on galaxy halo scales.
We summarize our results and conclude in Section 5. 


\section{Cosmological Sensitivity of Halo Merger Rates}
\label{sec:p2_analytic}

An analytic estimate for the halo merger rate was first derived by \cite{lacey.cole}, using the approach of \cite{Press.Schechter} to create the so-called `Extended Press-Schechter' (EPS) formalism. \cite{Sheth.Tormen.2002} derived a major correction to Press-Schechter theory, accounting for ellipsoidal collapse; this was subsequently included in EPS theory, e.g.~by \cite{Zhang.2008}. The halo merger rate has also been measured in $N$-body simulations, starting with \cite{Lacey1994}. Early work by \cite{Gottlober.2001}, for instance, studied the dependence on environment, while \cite{Fakhouri&Ma2008} and \cite{Fakhouri.2010} used the Millenium simulations to obtain accurate global merger rates, providing a framework to count mergers and compare numerical results to EPS predictions, as well as a universal fitting formula. We will use these results as the basis for most of our calculations. (For an alternative approach, that counts the rate per progenitor instead of the rate per descendant halo, see \citealt{genel.2009}.) 

Given the indirect connection between galaxies and haloes, the galaxy merger rate should behave slightly differently from the halo merger rate, as discussed further in Section 4. \cite{Stewart.2009} used $N$-body simulations to estimate how observable indicators of galaxy mergers should scale with galaxy luminosity, stellar mass, merger mass ratio and redshift. More recently, galaxy merger rates have been estimated from hydrodynamical simulations \citep[e.g][]{Sublink.Rodriguez_Gomez.2015}. We expect these estimates to be more accurate than earlier, dark-matter only results, although they are typically only applicable to a single cosmology.

\subsection{Merger Rate Definitions}
We will follow the definitions of \cite{Fakhouri&Ma2008} in describing the merger rate: first, the symmetric merger rate $B(M_1, M_2, z_D)dM_1dM_2$ is the average rate per unit volume per unit redshift, between two progenitors with masses in the ranges  $[M_1, M_1+dM_1]$ and $[M_2, M_2+dM_2]$ respectively, where $z_D$ is the redshift at which the descendant is identified. This rate has units of mergers/volume/(unit redshift).
If we want to express the rate in terms of the descendant mass $M_0 = M_1 + M_2$ and the merger mass ratio $\xi=M_2/M_1$ instead, we can use the function $B(M_0, \xi, z_D)dMd\xi$, also with units mergers/volume/(unit redshift). 
If we normalise the rate by the halo number density $n(M_0, z_D)$, we get the dimensionless quantity $B/n$, with units mergers/$dz/d\xi$:
\begin{equation}
    B/n \equiv \frac{B(M_0, \xi, z_D)}{n(M_0, z_D)}
\end{equation}
This quantity will be the basis for all the rates that we consider in this paper. We can also integrate $B/n$ in order to calculate the rate of all mergers between mass ratios x and X, at fixed descendant mass.
\begin{equation}
    \frac{dN}{dz}(M_0, z_D, x, X) = \int_x^X \frac{B}{n}(M_0, z_D) d\xi
\end{equation}

\subsection{Analytical Models}
The Extended Press-Schechter (EPS) formalism provides an analytic framework to estimate the merger or growth rates of dark matter haloes \citep{lacey.cole}, based on the spherical collapse model. The merger rate per halo, as a function of the descendant mass $M_0$ and the merger ratio $\xi = M_2/M_1$, is 
\begin{equation}
    \frac{B(M_0, \xi, z)}{n(M_0, z)} = \sqrt{\frac{2}{\pi}}\frac{d\delta_c}{dz}\frac{1}{\sigma(M_1)}\left|\frac{d \ln{\sigma}}{d \ln{M_1}}\right|\left[1 - \frac{S(M_0)}{S(M_1)}\right]^{-3/2}\,, 
\end{equation}
where $M_1 = M_0/(1+\xi)$ is one of the progenitors,\  $\delta_c$ is the critical overdensity for collapse, and $S \equiv \sigma^2(M)$ is the variance of the linear density field smoothed at a scale corresponding to a mass $M$. 

\begin{figure}
    \centering
    \includegraphics[width=0.9\linewidth]{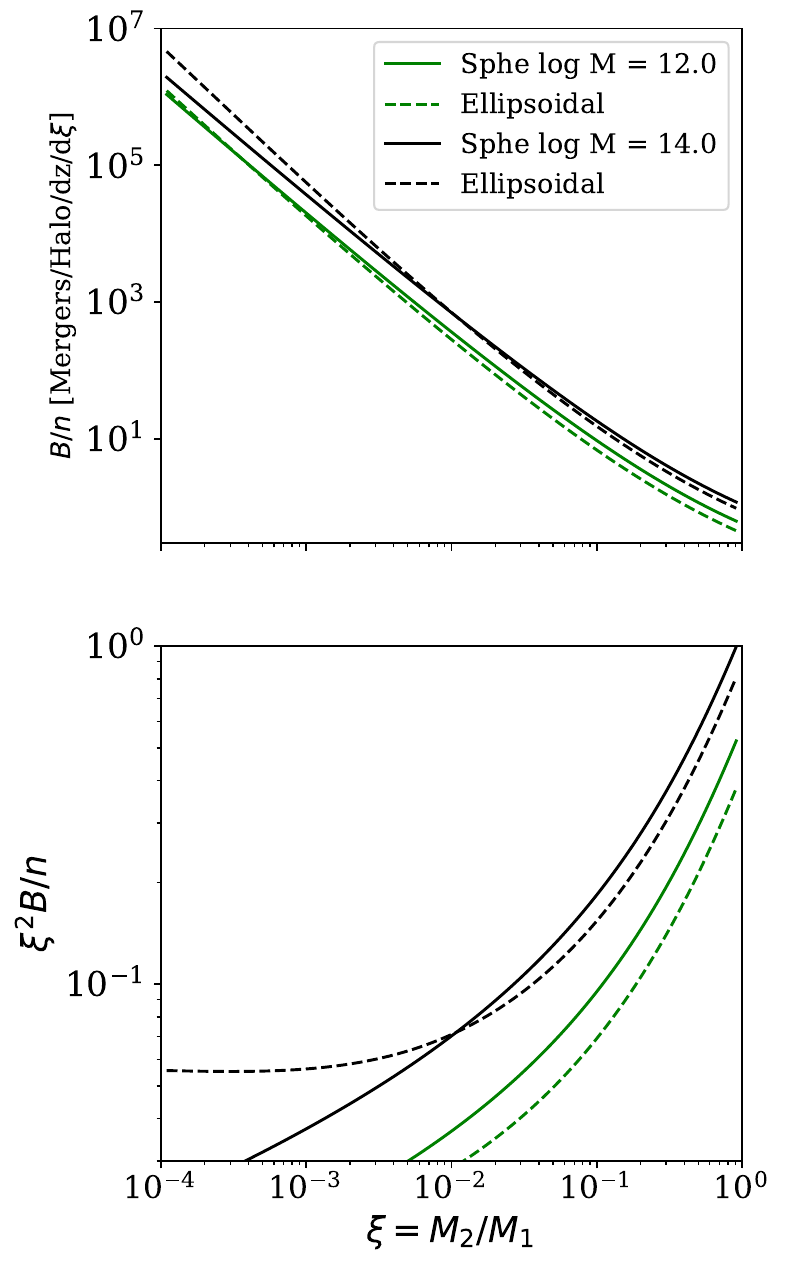}
    \caption{Halo merger rates predicted by the spherical collapse \citep{lacey.cole} and ellipsoidal collapse \citep{Zhang.2008} models, for the two halo masses indicated, at z=0.1. The bottom panel shows the rate weighted by $\xi^2$ to highlight the differences.}
    \label{fig:B/n = f(xi)}
\end{figure}

This model can be made more accurate by using an ellipsoidal collapse model instead \citep{Sheth.Tormen.1999, Sheth.Tormen.2002}. \cite{Zhang.2008} provide an ellipsoidal collapse correction to the spherical collapse merger rate 
\begin{equation}
\label{eq:zhang}
\begin{split}
    \frac{B(M_0, \xi, z)}{n(M_0, z)} &= \frac{B(M_0, \xi, z)}{n(M_0, z)}\bigg\rvert_{\rm sph} \\ 
    & \times A_0\exp\left(-\frac{A_1^2\Tilde{S}}{2}\right)\left[1+A_2\Tilde{S}^{3/2}\left(1+\frac{A_1\Tilde{S}^{1/2}}{\Gamma(3/2)}\right)\right]\,,
\end{split}
\end{equation}
where $A_0 = 0.8661(1-0.133\nu_0^{-0.615})$, $A_1 = 0.308\nu_0^{-0.115}$, $A_2 = 0.0373\nu_0^{-0.115}$, $\nu_0 = \omega_0^2/S(M_0)$, $\tilde S = \Delta S/S(M_0)$, $\Delta S = S(M_1) - S(M_0)$, and $\omega \equiv \delta_c(z)$.
The difference between spherical and ellipsoidal collapse-based merger rates is illustrated in Fig.~\ref{fig:B/n = f(xi)}. Throughout this paper, our analytical predictions will all be based on Eqn.~\ref{eq:zhang}.

\subsection{Cosmological Dependence of Merger Rate}
Analytical models provide a practical way to estimate how merger and growth rates will vary with the cosmological parameters. The merger rate is sensitive to cosmology through the power spectrum, as reflected by the variance of the smoothed field of density perturbations $S(M|\Omega_m, \sigma_8)$, and the growth factor $D$, through the critical overdensity for collapse $\delta_c(z) = \delta_c/D(z|\Omega_m)$ where $\delta_c = 1.686\Omega_m^{0.0055}$. Details of how $\Omega_m$ and $\sigma_8$ influence the cluster number count and formation time through their effect on the matter power spectrum and linear perturbation growth rate, as well as the resulting banana-shaped constraints, are discussed in \cite{Amoura2021}. 

\begin{figure}
    \centering
    \includegraphics[width=1\linewidth]{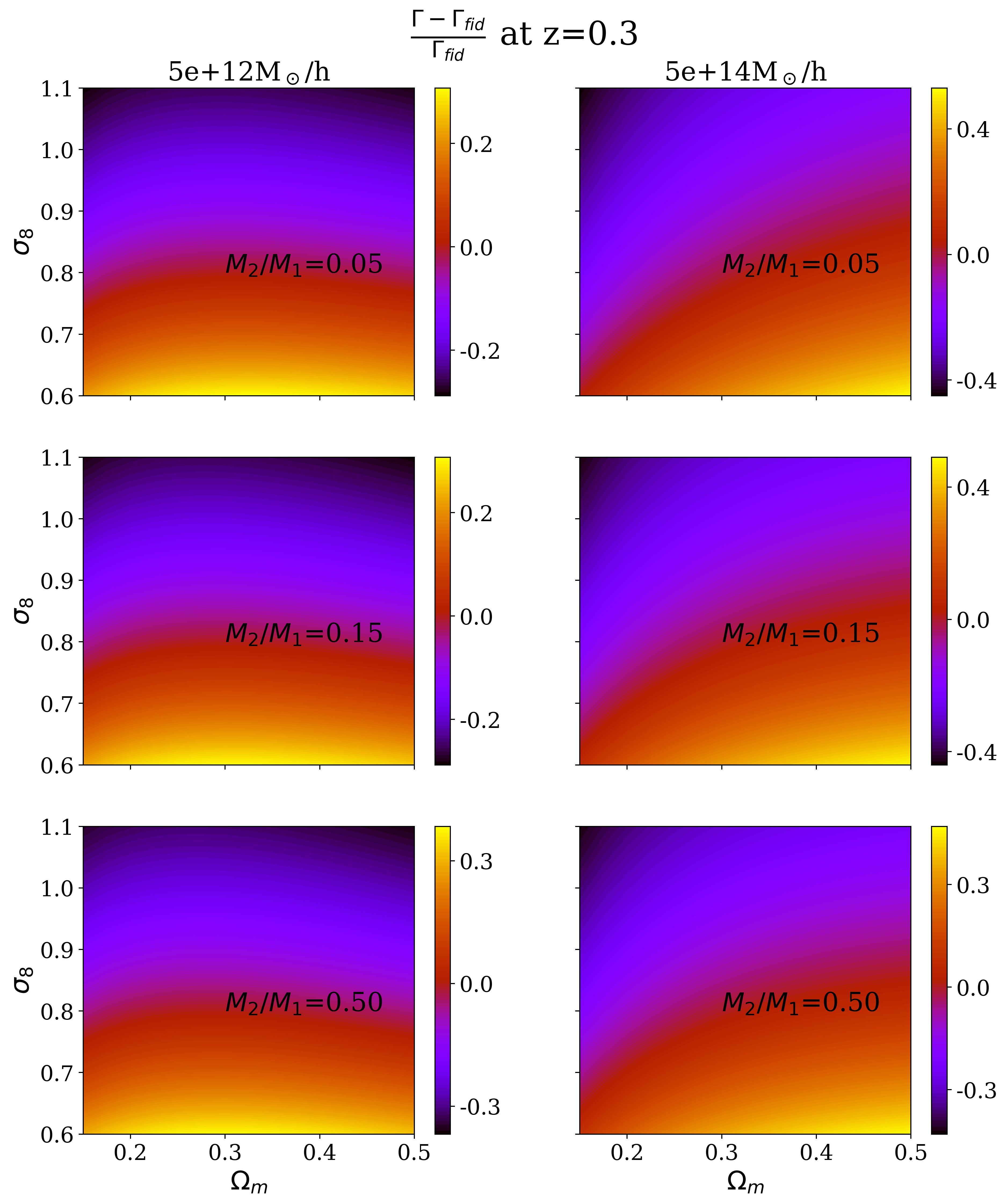} 
    \caption{Variation in the merger rate per halo $B/n$ at $z = 0.3$, as a function of $\Omega_m$ and $\sigma_8$, relative to a fiducial rate for $\Omega_m =0.3$ and $\sigma_8 = 0.8$. The rate is calculated assuming the ellipsoidal collapse model (Eqn.~\ref{eq:B/n ratio}).}
    \label{fig:B/n vs omega_m and sigma8}
\end{figure}

In Fig.~\ref{fig:B/n vs omega_m and sigma8}, we show how the merger rate estimated from Eqn.~\ref{eq:zhang} varies in the $\Omega_m$--$\sigma_8$ parameter space, for group- (left-hand panels) and cluster-mass (right-hand panels) haloes, and for three different mass ratios (top to bottom), at $z=0.3$. The colour scale shows the variation in the rate relative to a fiducial value calculated for $\Omega_m=0.3$, $\sigma_8=0.8$:  
\begin{equation}
    \Delta B/B_{fid} = \frac{B/n(M,z,\xi |\Omega_m, \sigma_8) - B/n(M,z,\xi |0.3, 0.8) }{B/n(M,z,\xi |0.3, 0.8)}\,.
\label{eq:B/n ratio}
\end{equation}
We see that the sensitivity to $\Omega_m$ and $\sigma_8$ is independent of merger mass ratio. The rate varies by about $\pm$20\%\ to $\pm$40\%\ over the range of $\sigma_8$ considered here, but depends only weakly on $\Omega_m$, with almost no dependence at the group mass scale. Thus, the halo merger rate can in principle be used to measure $\sigma_8$ independently from $\Omega_m$.

\subsection{Average Halo Growth}
\label{sec:AHG}
Material accreted onto a halo through mergers will settle into the main potential through tidal stripping and dynamical friction, over a timescale on the order of the dynamical time $t_{\text{dyn}}$. Thus, in addition to the instantaneous merger rate, we also consider the 
the net increase in halo mass over this timescale. For a given final redshift $z_0$, we first calculate the redshift $z_1$ corresponding to one dynamical time in the past.
The amount by which a halo grows over this redshift interval should then be
\begin{equation}
    \Delta M = \int^{z_1}_{z_0}dz\int_{0}^{1}B/n(M(z), z, \xi)\left[ \frac{\xi}{1+\xi}M(z)\right]d\xi\,.
\end{equation}
Since the merger rate varies slowly with mass, and the dynamical time is short enough that major mergers are rare, we can make the approximations
$B/n(M(z), z, \xi) \sim B/n(M_0, z, \xi)$ and $(1/(1+\xi))M(z) \sim M_1$ and define the Average Halo Growth (AHG) as:
\begin{equation}
    AHG(M_0, z_0) = \left<\frac{\Delta M}{M_1}\right>\equiv \int^{z_1}_{z_0}dz\int_{0}^{1}B/n(M_0, z, \xi)\xi d\xi\,, 
\end{equation}
that is, it is the increase in mass a halo experiences over one preceding dynamical time, relative to its initial mass, as a function of the final mass and redshift.

\begin{figure}
    \centering
    \includegraphics[width=1\linewidth]{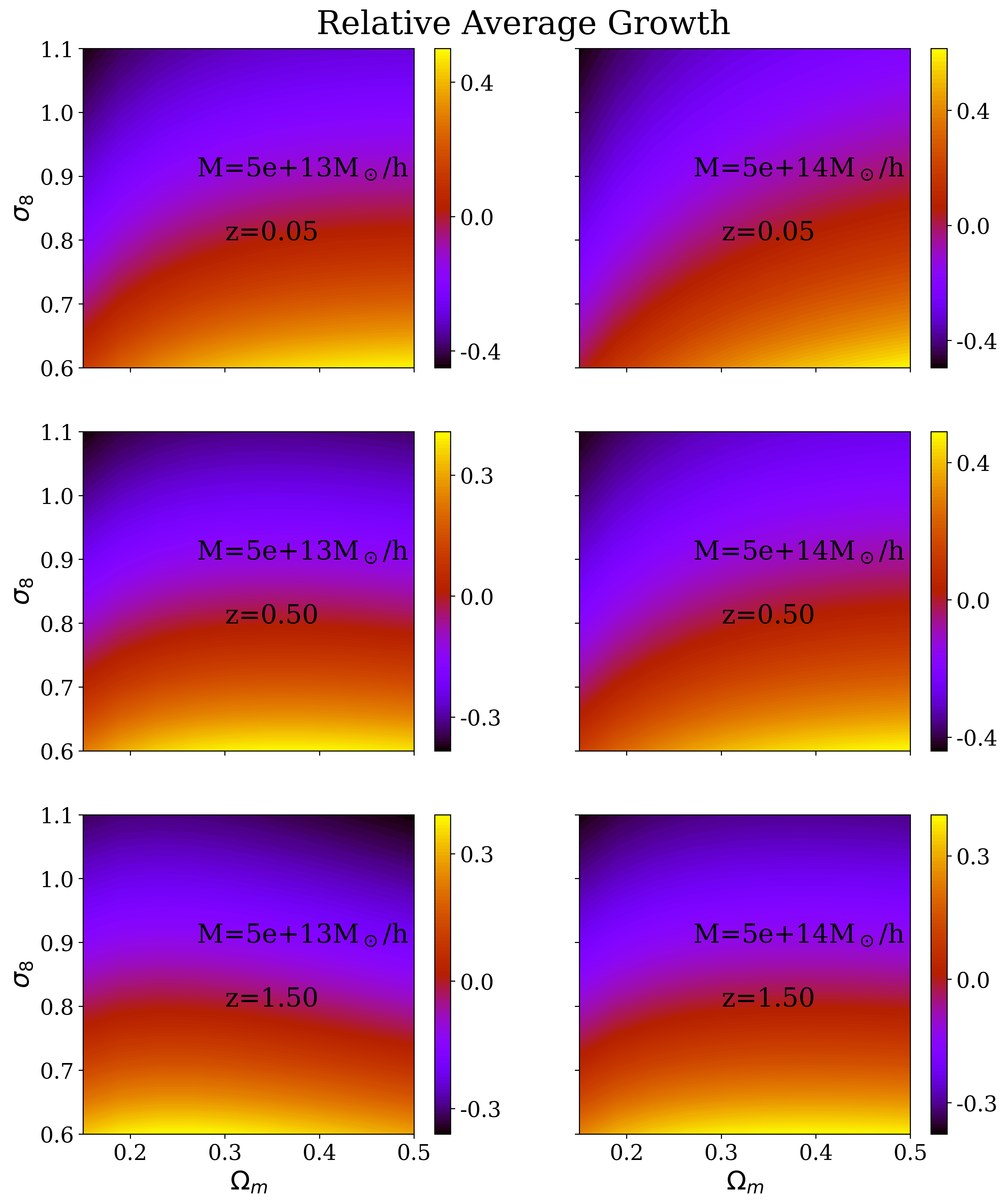}
    \caption{The average halo growth (AHG) over the last dynamical timescale $t_{\text{dyn}}$, as a function of $\Omega_m$ and $\sigma_8$ relative to the value at $\Omega_m=0.3$ and $\sigma_8=0.8$, for the masses and redshifts indicated.}
    \label{fig:average_growth_analytical}
\end{figure}

Fig.~\ref{fig:average_growth_analytical} shows the cosmological dependence of the AHG, for three different redshifts (top to bottom), and the same group and cluster masses as in Fig.~\ref{fig:B/n vs omega_m and sigma8}. As expected, haloes tend to grow faster at these redshifts in low $\sigma_8$ and/or high $\Omega_m$ universes. The influence of $\Omega_M$ becomes weaker relative to $\sigma_8$  for lower masses, and for higher redshifts. We note that contours of constant AHG are almost orthogonal to those typical of cluster abundance or weak lensing constraints (cf. Paper I). While the amplitude of the variation depends on the mass and redshift, there is generally between 50\% to 100\% change in the AHG between cosmologies with $\sigma_8 = 0.7$ and those with $\sigma_8 = 0.9$. 

\subsection{Large-growth Systems}
\label{sec:LGS}
Another summary statistic with a close connection to observable phenomena is the fraction of haloes that experience a large increase in mass over a given period of time. As for the AHG, we choose the dynamical time $t_{\text{dyn}}$ as the relevant timescale, and count the fraction of systems that have grown by more than 1/3 over this time. To estimate this fraction analytically, we make the approximation that the growth involves a single large merger with $\xi > 1/3$, such that
\begin{equation}
\begin{split}
    LGS(M_0, z_0) &\equiv \int^{z_1}_{z_0}dz\int_{1/3}^{1}B/n(M_0, z_0, \xi) d\xi\,.
\end{split}
\end{equation}

\begin{figure}
    \centering
    \includegraphics[width=1\linewidth]{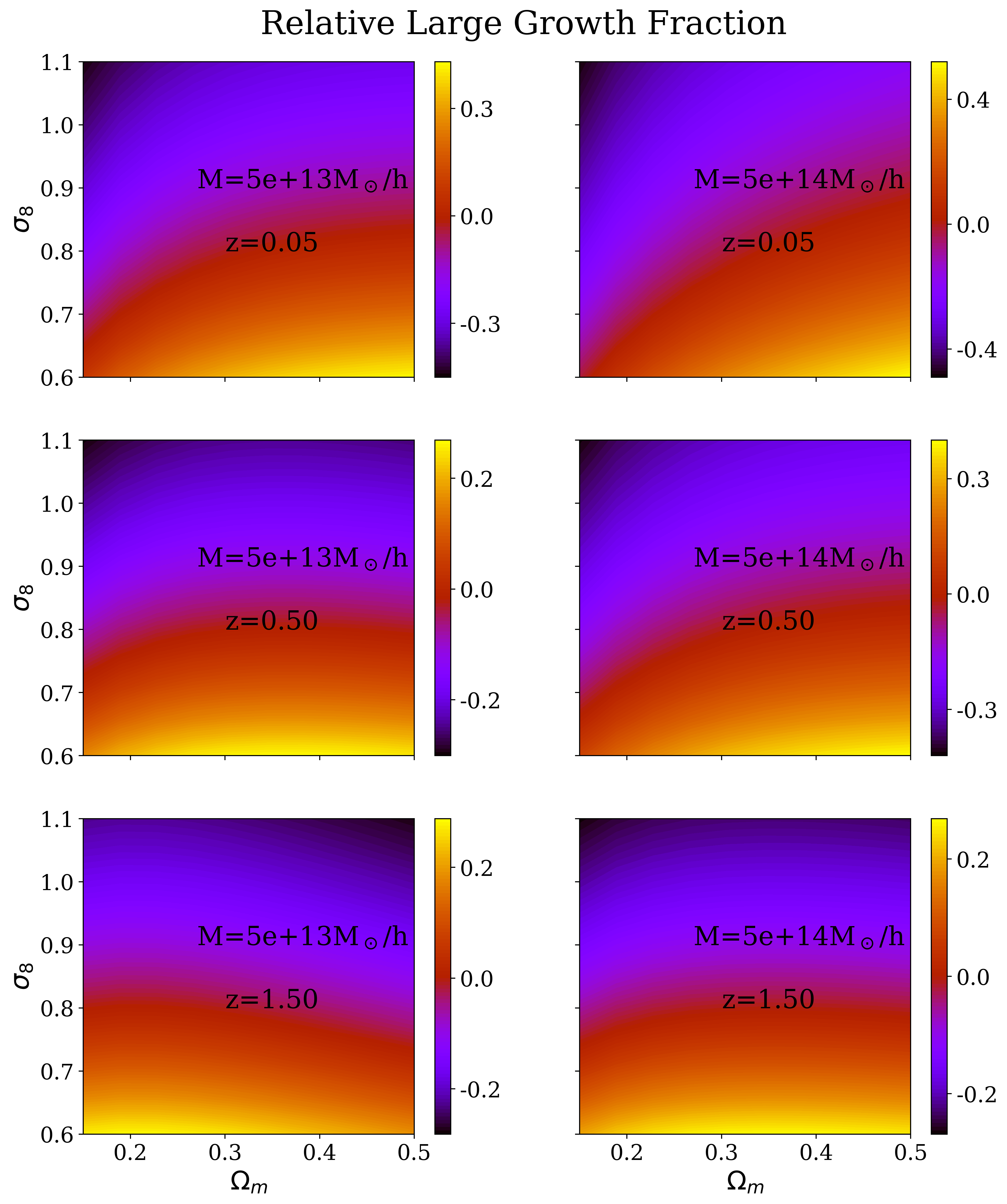}
    \caption{The fraction of large-growth systems (LGS) as a function of $\Omega_m$ and $\sigma_8$ relative to the value at $\Omega_m=0.3$ and $\sigma_8=0.8$, for the same mass and redshift bins as in Fig.~\ref{fig:average_growth_analytical}}.
    \label{fig:large_growth_analytical}
\end{figure}

Fig.~\ref{fig:large_growth_analytical} shows how
the LGS fraction depends on $\Omega_m$ and $\sigma_8$, for the same mass and redshift bins as Fig.~ \ref{fig:average_growth_analytical}. The cosmological dependence is almost identical to that of the AHG, such that both quantities could in principle provide cosmological tests of comparable sensitivity. 

\section{Comparison to simulations} 
\label{sec:p2_simulations}

As discussed in Paper I, the analytic models of the previous section are only approximate. To test their validity, we will also consider merger rates measured in several different $N$-body simulations.

\subsection{Simulation Data}
We use a set of dark-matter-only simulations to test the analytical predictions of the ellipsoidal collapse model. These include publicly available halo catalogues and merger trees, but also our own set of simulations run for different cosmologies, as follows : 
\begin{enumerate}
    \item The Illustris-TNG simulation \citep{nelson2019illustristng} which uses the {\sc Subfind} halo finder and the {\sc Sublink} merger tree algorithm \citep{Sublink.Rodriguez_Gomez.2015}. 
    \item The Bolshoi/BolshoiP simulation \citep{Bolshoi.Klypin_2011}, with a halo catalog generated with {\sc Rockstar} \citep{Rockstar.Behroozi_2012} and merger trees generated with the {\sc Consistent Trees} algorithm \citep{Consistent_trees.Behroozi_2012}.
    \item A set of 9 of our own simulations, introduced in Paper I. We will refer to these as MxSy, where x can be 25/3/35 for $\Omega_m =0.25/0.3/0.35$ respectively, and y can be 7/8/9, for $\sigma_8=0.7/0.8/0.9$ respectively. These simulations were run with {\sc Gadget 2} \citep{Gadget2}, and the halo catalogue and merger trees were generated with the {\sc Amiga Halo Finder} \citep[AHF;][]{AHF}. 
\end{enumerate}
Simulation parameters are summarized in Table \ref{table:p2_sims}.

\begin{table*}
\begin{tabular}{ |c|c|c|c|c|c|c|c| } 
 \hline
 Simulation Name & $\Omega_{\rm m}$ & $\sigma_8$ & $m_{\text{part}}$ [$M_\odot /h$]& $N_{\rm part}$ & merger tree & $N_{\rm snap}$  \\ 
 \hline
Illustris TNG & 0.31 & 0.81 & $3 \times 10^9$ & $625^3$ & Sublink & 100 \\ 
 Bolshoi & 0.27 & 0.82 & $1.35 \times 10^8$ & $2048^3$ & Consistent Trees & 181 \\
 BolshoiP & 0.31 & 0.82 & $1.55 \times 10^8$ & $2048^3$ & Consistent Trees & 178  \\
 MxSy & 0.25/0.3/0.35 & 0.7/0.8/0.9 & $4 \times 10^9$ & $512^3$ & Amiga Halo Finder & 44 \\
 \hline
\end{tabular}
\caption{Summary of the simulations used and their main parameters, including the cosmological parameters, the particle mass, the total number of particles $N_{\rm part}$, the merger tree code, and the number of snapshots $N_{\rm snap}$ used to make the merger trees. The MxSy simulations are a set of 9 of our own simulations that span a range of different values of $\Omega_{\rm m}$ and $\sigma_8$.}
\label{table:p2_sims}
\end{table*}

\subsection{Merger Rates}

To estimate the merger rate per descendent halo, $B/n$, in the simulations, we count the individual mergers associated with a given descendent as follows. Going to the previous snapshot, we identify all $N_{\text{prog}}$ progenitors of the descendent, and count a total of $N_{\text{prog}}-1$ mergers, each with the most massive progenitor (implying merger mass ratios $\xi < 1$). The exact definition of the progenitors varies, depending on the simulation and the merger tree algorithm. For AHF, the merger tree is constructed by correlating particles from a given descendant halo to haloes from an earlier snapshot. Because the AHF halo finder is inclusive, i.e. particles from subhalos are also part of the main halo, mergers are counted as soon as a progenitor's particles are included in a descendent. This occurs when they are enclosed in a spherical region of density 200 times the critical density; for minor mergers between spherical halos, this corresponds roughly to the moment when the virial radii of the two halos first overlap. 

We note that AHF does not always recognize events where halos merge, pass through each other, and then merge again, counting each event as a separate merger.

The Bolshoi simulations use similar merger criteria and a similar mass definition. By comparison, the {\sc Subfind} groupfinder used in the Illustris simulations groups particles into halos using a friends-of-friends criterion, together with dynamical information; also its halo masses are {\it exclusive} of substructure.

In addition, each simulation has a different snapshot frequency, which can affect the measured merger rate, especially at high redshift. After some experimentation, for most of our tests we restricted ourselves to merger rate estimates from our own simulations, where the analysis is homogeneous. 

\begin{figure*}
    \centering
    \includegraphics[width=0.95\linewidth]{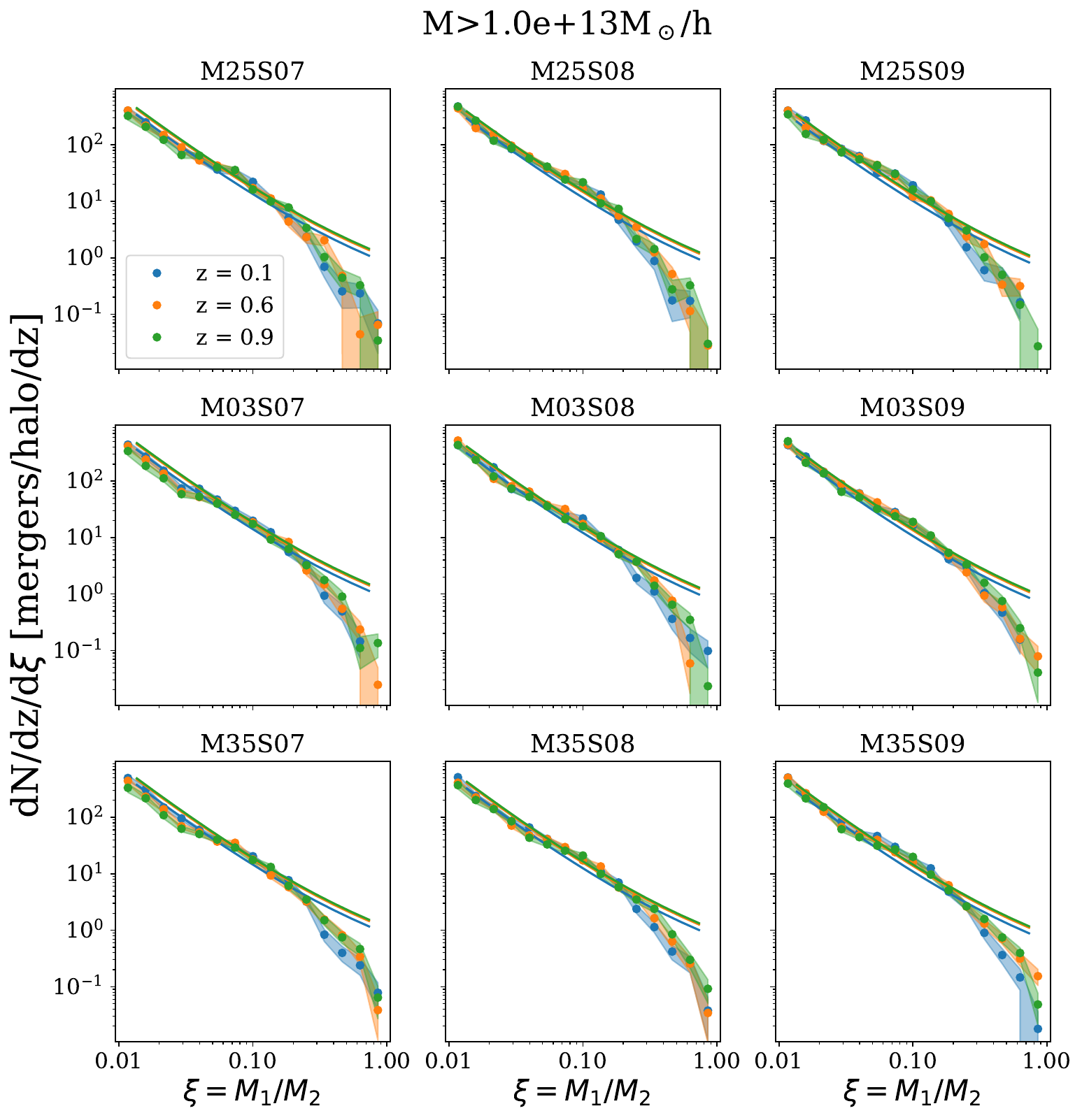}
    \caption{The merger rate per descendant halo as a function of mass ratio $\xi$ for each of the MxSy simulations (points and shaded regions), compared to the EPS rate predicted by the ellipsoidal collapse model. Note the deficit of major mergers, relative to the analytic predictions.}
    \label{fig:merger_rate_sims}
\end{figure*}

Fig.~\ref{fig:merger_rate_sims} shows the numerical merger rate as a function of mass ratio $\xi$ from each simulation, compared to the analytical predictions. Generally, the numerical results are in reasonable agreement with the analytical models, but include far fewer major mergers. This may be an artefact of our method for counting mergers. If the progenitors of a given descendent include one large halo and several smaller ones, we always count $N-1$ minor mergers between the largest progenitor and each of the other progenitors. With a higher frequency of snapshots, we might find that intervening major mergers had occurred between pairs of low-mass progenitors, before they merged with the largest progenitor. In addition, tidal stripping can cause systems to lose some mass even before they are recorded as merging. Either of these effects could explain the deficit of large mass-ratio mergers and the slight excess of lower mass-ratio mergers.

In order to study the cosmological dependence of the merger rate more specifically, we calculated the number of mergers between z=0.05 and z=0.45, for mass ratios between $ 0.01<\xi<0.03$, $0.03<\xi<0.1$ and $0.1<\xi<0.3$. We avoided major mergers, given the potential problems described above. We also restricted ourselves to lower redshifts, where the snapshot cadence is reasonably frequent relative to the dynamical time. We considered all haloes with $M > 10^{13}M_\odot $, first to avoid mass resolution effects at low mass, and second to have enough statistics given the first constraint. Binning all masses together is reasonable, given that the merger rate is only weakly sensitive to mass, going as $\sim M^{0.13}$ \citep{Fakhouri.2010, genel.2009}.  

\begin{figure*}
    \centering

    \includegraphics[width=0.95\linewidth]{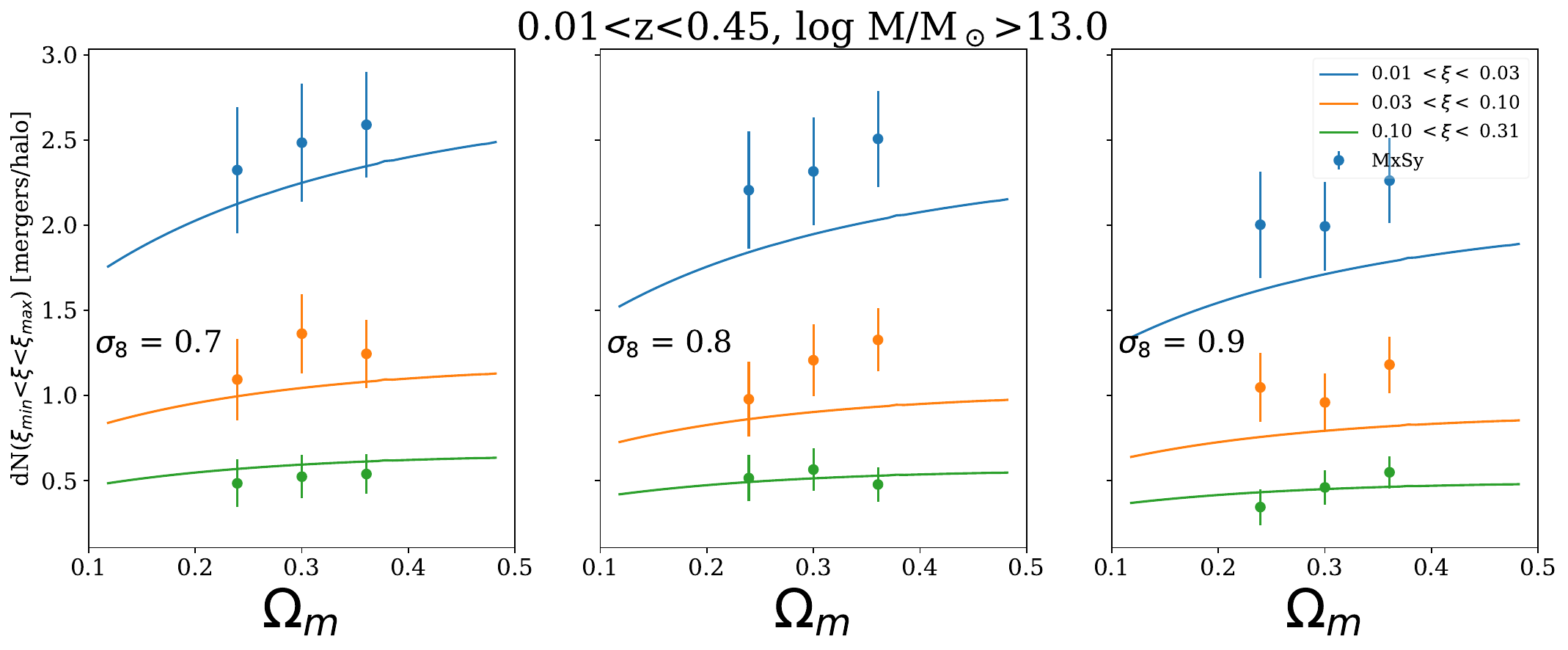} \\ 
    \includegraphics[width=0.95\linewidth]{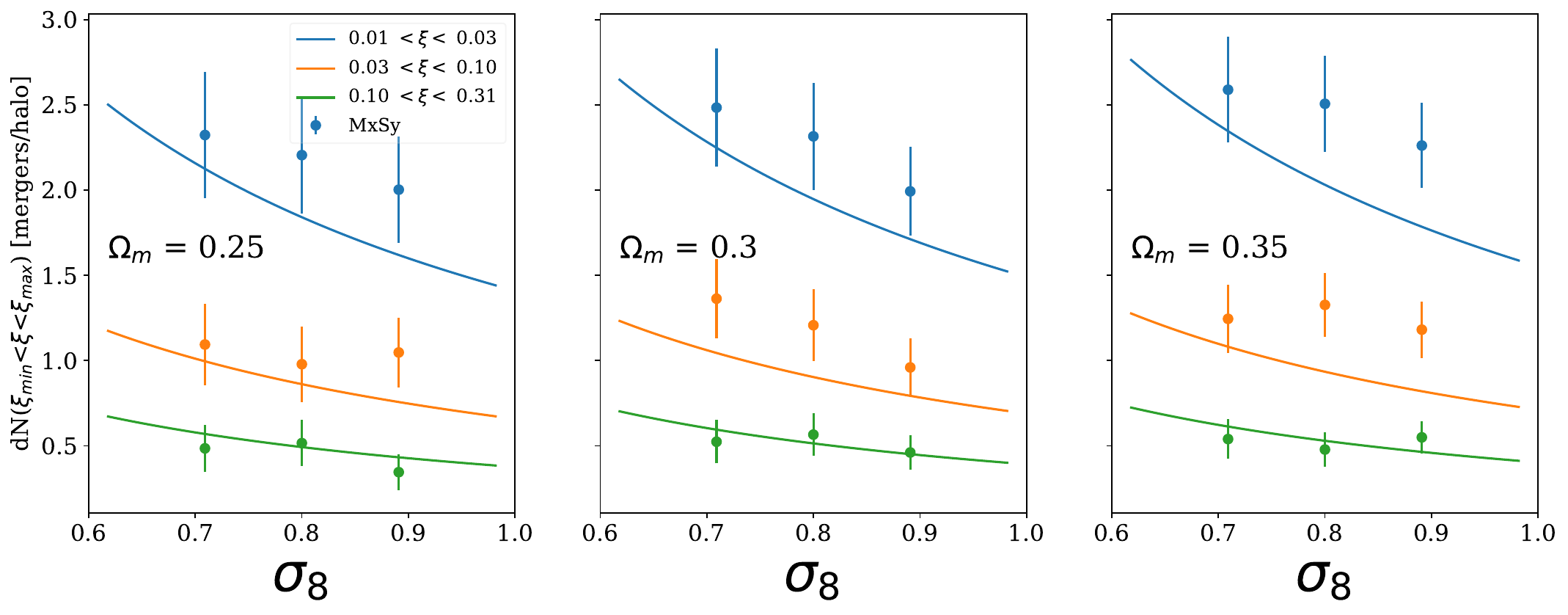}
    \caption{Cosmological dependence of the merger rate for various mass ratios. The points with errorbars indicate rates estimated from the MxSy simulations, while the smooth curves show the analytic predictions.}
    \label{fig:merger_rate_cosmo}
\end{figure*}

Fig.~\ref{fig:merger_rate_cosmo} shows these merger rates, as a function of $\Omega_m$ at fixed $\sigma_8$ (top panels), and as a function of $\sigma_8$ at fixed $\Omega_m$ (bottom panels), compared to the analytical predictions. Both numerical and analytic results show the same general behaviour. The simulations contain more minor mergers ($0.01 <\xi<0.03$) than predicted by theory, which may reflect the counting problems discussed above, but the dependence on cosmological parameters is similar between the numerical and analytic results.

\subsection{Average Halo Growth}

\begin{figure*}
    \centering
    \includegraphics[width=0.95\linewidth]{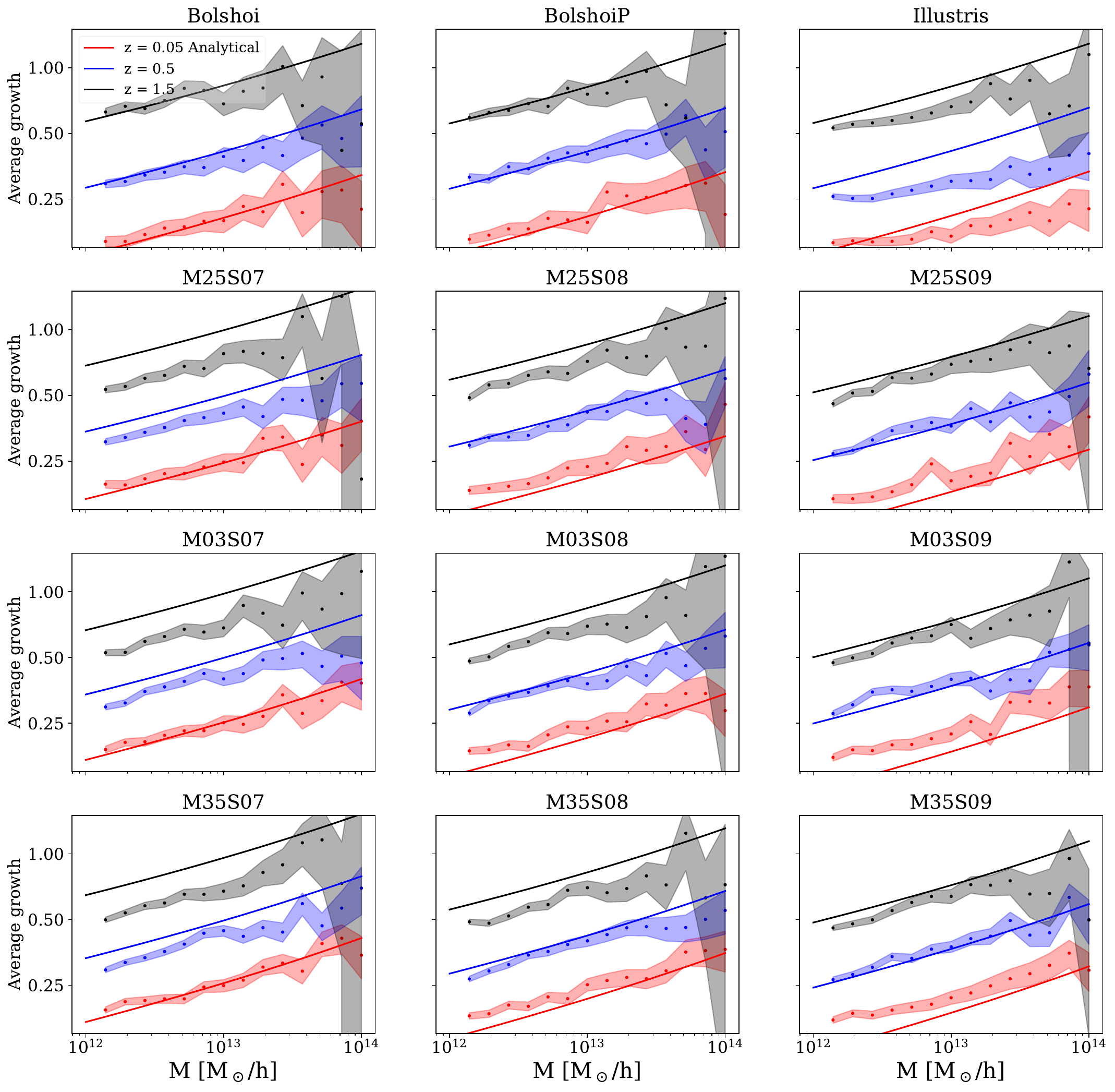}
     \caption{The average growth of haloes since the last dynamical timescale $t_{dyn}$, measured in the simulations indicated. The shaded areas represent Poissonian errors. Solid lines show the EPS predictions for comparison.}
    \label{fig:av_growth_sims}
\end{figure*}

Next, we compare the average halo growth rate measured in simulations to the rate predicted by EPS theory. The comparison is particularly interesting, since the simulations and halo finders have finite resolution, and will always miss a component of the merger history below their resolution limit. For this test, we consider results from all the simulations listed in Table 1, to highlight the differences between them.  
For every halo in each simulation, we measure the mass growth over one dynamical time $\left(M(z-z_{dyn})-M(z)\right)/M(z)$
and average this quantity in each mass bin. We then calculate the same quantity in analytical models by integrating the instantaneous merger rate over the same redshift range. The resulting rates are shown in Fig.~\ref{fig:av_growth_sims}. 

The Bolshoi simulations agree with analytical models well at all redshifts, while Illustris shows a flattening at high mass. The set of MxSy simulations agree well at low z and less at high z. Most simulations have lower growth rates at high redshift than predicted by theory. On the other hand, all the numerical results agree with the analytic predictions in the general mass and redshift dependence in the growth rate, often differing by a single overall shift in normalization. We speculate that halo-finding algorithms may be at the origin of this discrepancy between different simulations, and between simulations and analytical predictions, as discussed in the literature \citep{Sussing_Knebe_2011, avila.2014.sussing, hopkins.2010}. Cases of haloes losing mass, flyby events and other numerical artifacts introduced by the different ways haloes are defined, detected and linked in different halo-finder and merger tree algorithms can cause an artificial increase in the average growth. Even after accounting for and removing the most spectacular events, where haloes appear to gain several times their mass between consecutive snapshots, the average growth at low redshift remains larger than analytical predictions. 

We note that the agreement with theory is closest for the Bolshoi simulations, which also have the highest snapshot cadence; by comparison, the cadence is lowest for the MxSy simulations, so this may also account for some of the differences, notably the higher growth rates at low redshift, seen in Fig.~\ref{fig:av_growth_sims}. We have run and analyzed similar simulations with higher cadence, however, and found the trends with cadence are subtle and hard to establish conclusively. Finally, the EPS prediction itself may be inaccurate beyond some point, particularly over short timesteps \citep[e.g.][]{Sheth2004}.

We now consider the cosmological dependence of the AHG. Given the differences between different simulations and analysis tools shown above, we will restrict ourselves to our own MxSy simulations, which represent a homogeneous set. To simplify the comparison between simulations, we fit the simulation results with a power-law
\begin{equation}
\label{eq:power-law_fit}
    AHG(M) = A(M/M_0)^\alpha + p\,,
\end{equation}
where the parameters $A$, $M_0$ and $\alpha$ are fixed in each panel, and the normalisation $p$ varies with $\sigma_8$. For z=1, we find that a broken power law is a better fit. This fit is meant to reproduce the overall mass dependence of the AHG specifically for our set of simulations. We show the fits and discuss them further in Appendix \ref{sec:appendix1}.

In Fig.~\ref{fig:av_growth_sims_vs_analytical_s8}, we compare the dependence of the AHG on $\sigma_8$, for different values of $\Omega_m$, and for different masses and redshifts. As we have seen already in Fig.~\ref{fig:av_growth_sims}, the halo growth from simulations is significantly lower at high redshift than the analytic prediction. Accounting for this redshift-dependent offset, the numerical results show the predicted drop in growth with increasing $\sigma_8$, but seem less sensitive to $\sigma_8$ than expected, particularly at high redshift. The numerical values in high-$\sigma_8$ (low growth) cosmologies exceed the analytic predictions. The origin of this discrepancy is not immediately clear. One possibility is that because of the relatively large spacing between snapshots in the MxSy simulations, the true growth rate is {\it over}-estimated in cosmologies where it is intrinsically low. We will investigate this possibility in future work.

\begin{figure*}
    \centering
    \includegraphics[width=0.9\linewidth]{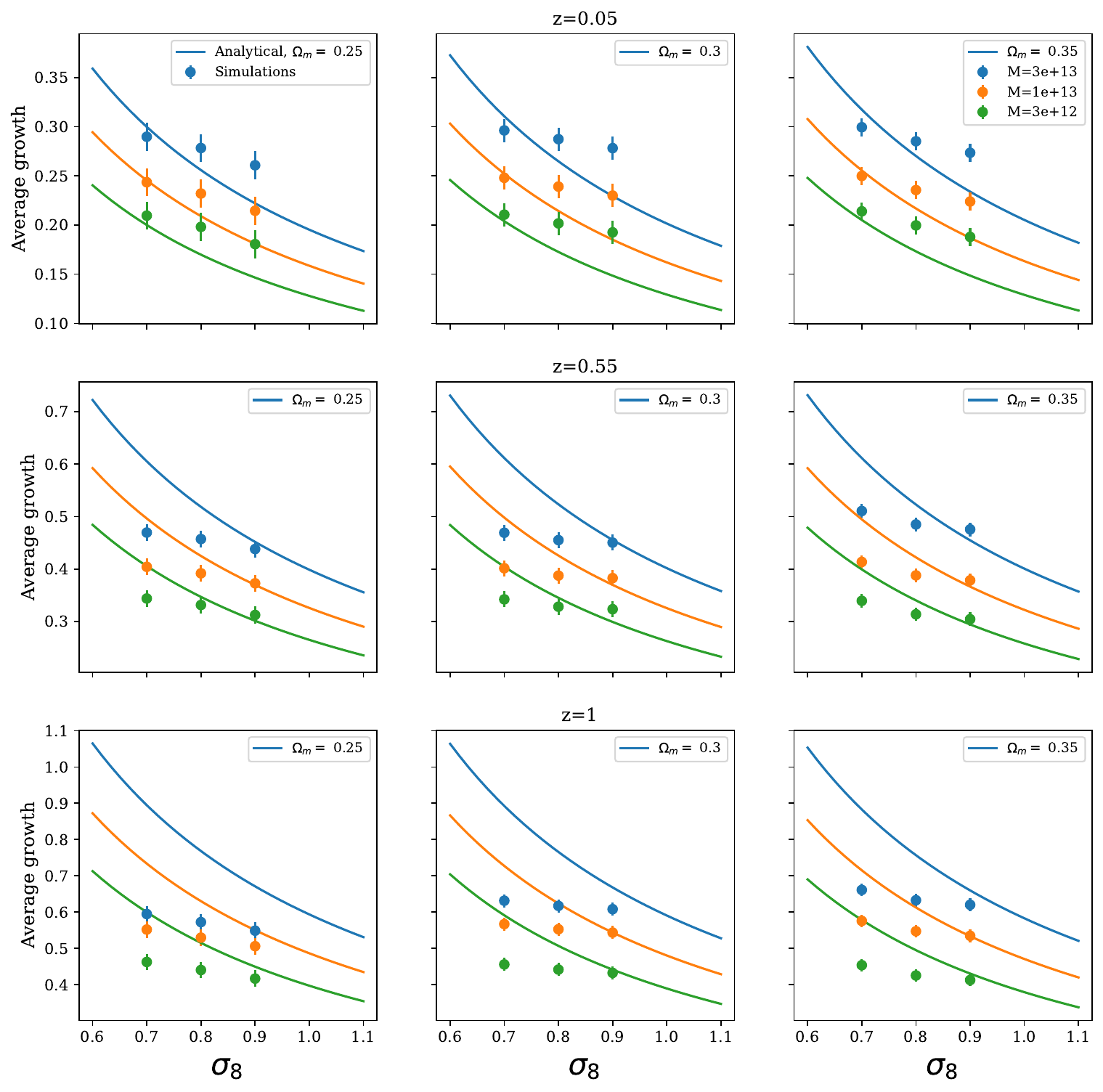}
    \caption{Average halo growth as a function of $\sigma_8$, for the redshifts, masses, and values of $\Omega_{\rm m}$ indicated. The points show the results measured in the MxSy simulations, while the curves show the EPS prediction.}
    \label{fig:av_growth_sims_vs_analytical_s8}
\end{figure*}

\subsection{LGS Fraction}

\begin{figure*}
    \centering
    \includegraphics[width=0.98\linewidth]{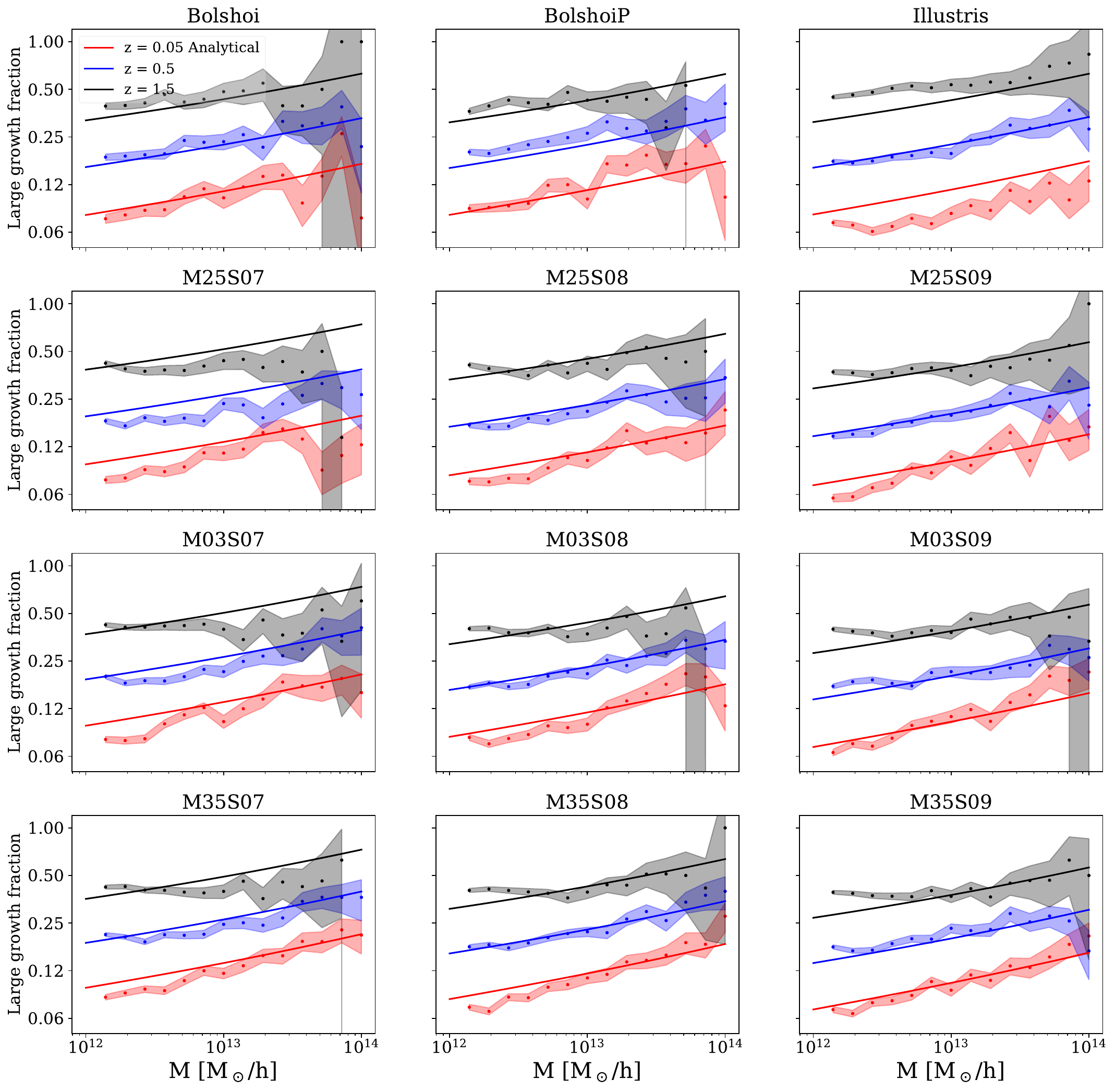}
     \caption{Fraction of haloes that experienced a large growth (>1/3) since the last dynamical timescale $t_{dyn}$, as a function of mass (points with shaded error regions). Solid lines show the EPS predictions.}
    \label{fig:large_growth_sims}
\end{figure*}

In Fig.~\ref{fig:large_growth_sims} we show the LGS fractions measured in the simulations, compared to the analytic predictions.
Unlike the AHG, all simulations are in agreement with the analytic predictions at z=0.05; at higher redshift, the mass dependence seems slightly flatter than predicted. Note that Illustris is the outlier again, finding considerably more large growth systems at high redshift.

\begin{figure*}
    \centering
    \includegraphics[width=0.9\linewidth]{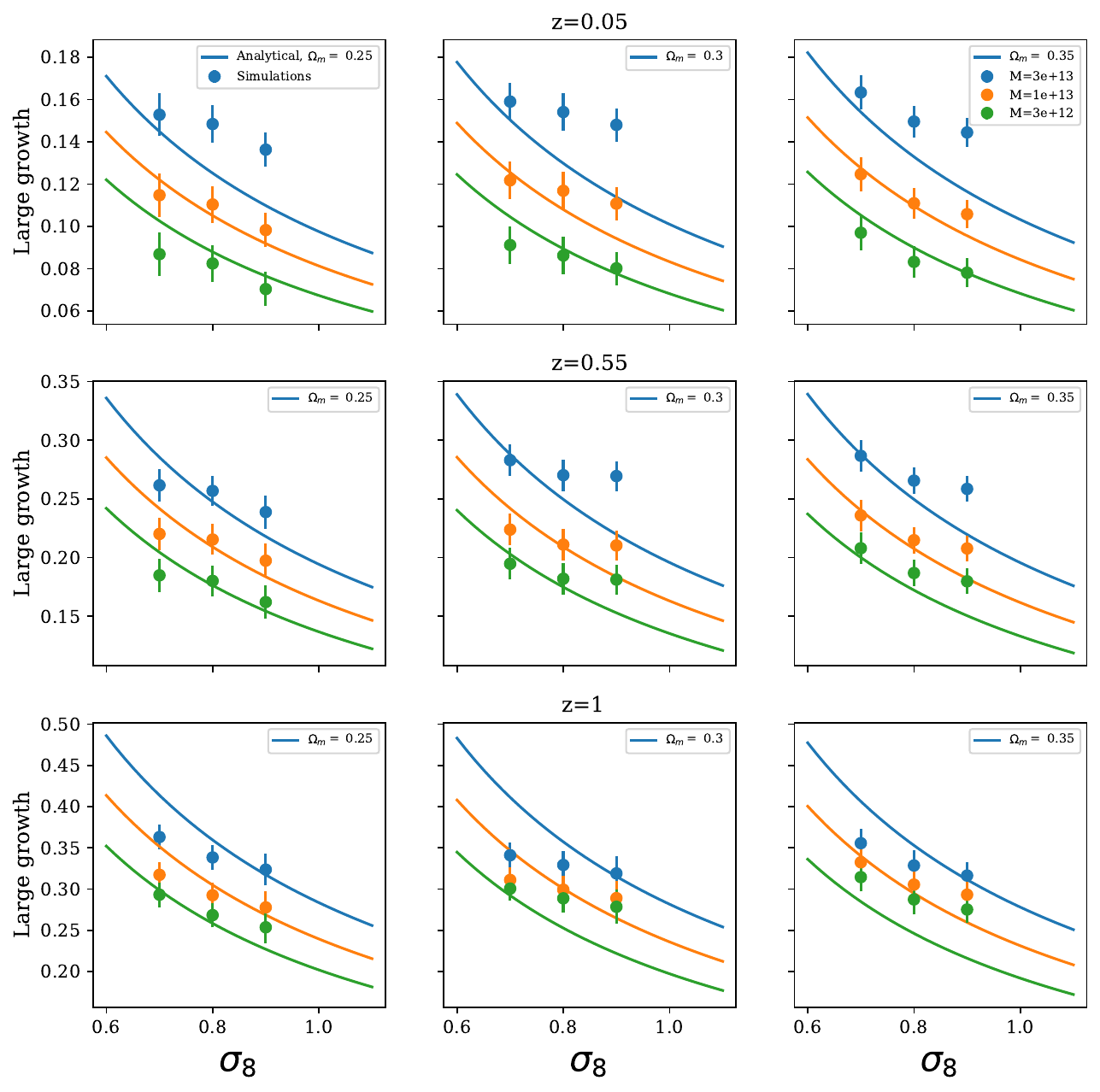}
    \caption{Comparison of the $\sigma_8$ dependence of the fraction of haloes with large growth between simulations and analytical models. The cosmological trend is similar, with a lower amplitude. }
    \label{fig:large_growth_sims_vs_analytical_s8}
\end{figure*}

As for the AHG, we fit the power law in Eqn.~\ref{eq:power-law_fit}, for each of the MxSy simulations, and show the variation of the normalisation at different values of $\sigma_8$ in Fig.~\ref{fig:large_growth_sims_vs_analytical_s8}. The range of variation of the LGS fraction with $\sigma_8$ is closer to the analytical predictions at low redshift, but still slightly smaller at higher redshift.

\subsection{Simulations vs.~Analytic Predictions: Summary}
Comparing numerical and analytic results, we find broad agreement, but also some discrepancies. Unfortunately, without further detailed work, it is not clear which if either is the most accurate, although we suspect at least some of the discrepancies are related to the snapshot cadence and merger tree algorithms used to analyse the simulations. The predicted and measured cosmological dependence are in closest agreement for the halo merger rate and the LGS fraction. Focussing on these quantities, we infer that we could differentiate between values of $\sigma_8$ between 0.7 and 0.9 if we could measure either one with a precision of better than 10\% while avoiding any observational systematics. In Section \ref{sec:observations} below, we will consider whether this goal is realistically achievable. 

\section{Observational prospects} 
\label{sec:observations}

The results of the previous section suggest that $\sim$10\% precision would be required in merger or growth rate measurements, in order to provide useful cosmological constraints. We will now examine whether this precision could be reached in practice. We consider tests on two scales, either the galaxy cluster scale, or the scale of individual galaxy haloes.

\subsection{Measuring Merger and Growth Rates on Cluster Scales}

While galaxy clusters are relatively rare, in the near future multiple missions and surveys including {\it Euclid} in the optical and IR \citep{sartoris.2016}, \emph{eROSITA} \citep{Pillepich2012} in the X-ray, CMB-S4 \citep{CMBS4} in the mm, and the ground-based UNIONS \citep{UNIONS2020}, DESI \citep{DESI2016}, and Rubin LSST \citep{LSST2009} surveys should produce mass-limited samples of $O(10^4)$ clusters with sufficient signal-to-noise ratio (SNR) to allow structural measurements. Clusters with sufficient SNR to detect major mergers should number in the hundreds or thousands. Furthermore, these will typically be low-redshift, massive systems where complementary information from many modalities is available, including galaxies with measured redshifts, weak and/or strong lensing mass models, X-ray surface brightness maps, and SZ maps in the sub-mm.

To measure the instantaneous halo merger rate for clusters would require identifying all infalling groups at or near the virial radius. This could be challenging due to projection effects and/or limited galaxy redshift information. Furthermore, one would need to estimate total masses for the infalling systems, with errors in the mean mass for a sample not exceeding 10\%. A realistic survey of $O(400)$ massive clusters with weak lensing mass maps might identify infalling systems in, say the range $\xi =$ 0.1-0.2, with 50-70\%\ completeness, over a narrow redshift range where the average number of mergers is one per cluster. The Poisson uncertainty in the merger rate would then be $1/\sqrt{400} = $ 5\%; the uncertainty in the mean mass of the infalling systems would be 200\%/$\sqrt{400} = $10\%, while the uncertainty in the completeness might be $\sim$20\%. We conclude that while the first two sources of uncertainty are close to the goal of 10\% errors, the uncertainty in the completeness would be too large to obtain useful cosmological constraints.

Alternately, one could consider measuring the LGS fraction. Systems that have recently accreted a third or more of their material would be easier to identify, via kinematic substructure, offsets between the gaseous, stellar and dark components \citep[e.g.][]{Clowe.2006, Mann.2012, Zenteno.2020}, or overall X-ray morphology \cite[e.g.][]{Yuan.2022}. Assuming these features can be detected regardless of projection effects, we may assume approximately ~100\% completeness in the LGS sample. Assuming a LGS fraction 20-30\% for massive clusters at low redshift, a sample of 400 might produce ~100 LGS systems, resulting in Poisson errors with the required uncertainty of 10\%. On the other hand, distinguishing between degrees of relaxation (e.g.~systems that had experienced large growth within the past 1.0 dynamical times, versus 2.0 or 0.5 dynamical times) might be more challenging, and would require extensive calibration with simulations.

Overall, we conclude that measuring the halo merger rate or growth rate on cluster scales seems challenging, but not impossible. At a minimum, future cluster samples should provide a consistency test for parameters derived from other methods.

\subsection{Measuring Merger and Growth Rates on Galaxy Scales}

Given that halo merger and growth rates depend only weakly on halo mass, and galaxy haloes are far more abundant, it is worth considering tests based on this smaller mass scale. Galaxy merger rates have been studied extensively, both theoretically, either through semi-analytical/semi-empirical models \citep[e.g.][]{Stewart.2009, Husko.2022} or hydrodynamical simulations \citep[e.g.][]{Sublink.Rodriguez_Gomez.2015, Pfister.2020, Contreras-Santos.2022}, and observationally \citep[e.g.][]{Lotz.2011, Xu.2012, Mundy.2017}. Two important complications arise in relating galaxy merger rates to galaxy halo merger rates; first, the delay between the two, and second, the relation between halo mass and stellar mass. We consider each of these in turn.

\subsubsection{Delay Time Due to Infall}

Mergers between galaxy haloes, as defined in most analyses of numerical simulations, occur around the virial radius. Assuming both haloes contain visible galaxies, halo mergers then lead to galaxy mergers, after some delay for infall to the centre of the main halo. Galaxy mergers are identified observationally using features---close pairs, tidal distortion and debris, or triggered starbursts---that trace the first and second pericentric passages. We note that the delay due to infall means that `merging' galaxies identified via these features at one redshift actually trace the halo merger rate at a higher redshift.

\begin{figure}
    \centering
    \includegraphics[width=1\linewidth]{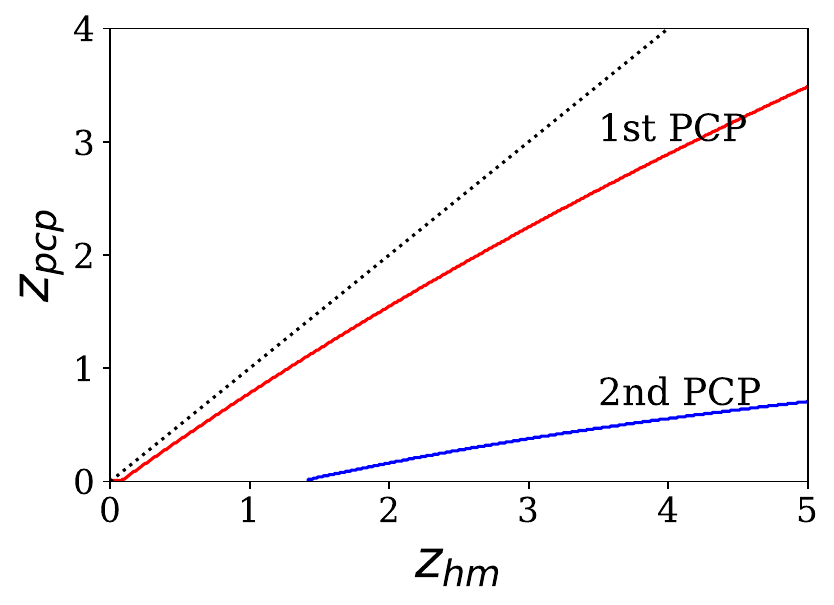}
    \caption{Redshift at which first (top curve) and second (bottom curve) pericentric passages would occur, assuming a halo merger at $z_{hm}$ and conservation of orbital properties. The dotted line shows a 1-1 correspondence for reference.}
    \label{fig:zpcp}
\end{figure}

Fig.~\ref{fig:zpcp} shows the observed redshift at which the first (top curve) and second (bottom curve) pericentric passages occur for a given halo merger redshift $z_{hm}$.
(The dotted line shows a 1--1 correspondence for reference.)
To calculate these, we have assumed that the orbital properties of the satellite and the potential of the main system are conserved, and that pericentric passages occur around 1/8 and 9/8 of the radial orbital period at the virial radius \citep{TB04}, which corresponds to 0.1 and 0.9 times the period of a circular orbit at the virial radius, $P_{\rm vir}$, or 0.06 and 0.57 times the Hubble timescale $H(z)^{-1}$ at the redshift $z_{hm}$.

From this figure, we see that while the first pericentric passage occurs at only slightly lower redshift than the initial merger, the second pericentric passage occurs significantly later, and is only observable for halo mergers at $z_{\rm hm} > 1.5$.
These calculations assume conservation of the orbit and the potential over 1 or more radial periods; the reality in major mergers is more complicated, and a significant fraction of orbits may get scattered in these cases (de Luna et al. in prep.).

\subsubsection{The Impact of the Stellar-to-Halo Mass Relation}

In this paper, we have considered the growth and merger rates for haloes. In contrast, observational studies of galaxy-scale mergers measure these rates as a function of luminosity or stellar mass. The stellar-to-halo mass relation (SHMR) is fairly well constrained from a variety of observations \citep[e.g.][and references therein]{Behroozi2019}, and has the form of a broken power-law that changes slope abruptly on group scales. Since the halo merger rate is close to a single power-law in halo mass ratio $\xi$, the shape of the SHMR should produce a kink in the merger rate measured as a function of {\it stellar} mass ratio. 

\begin{figure}
    \centering
    \includegraphics[width=1\linewidth]{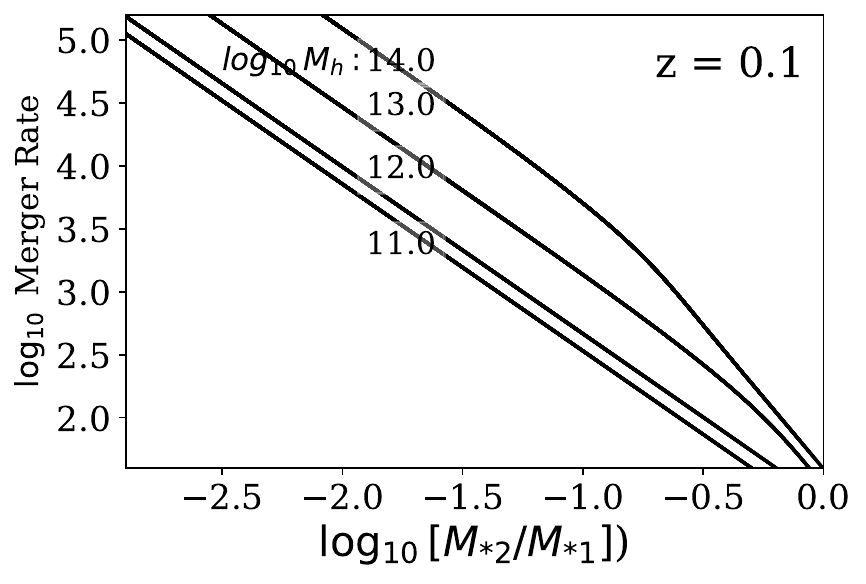}
    \caption{Merger rate onto central galaxies in haloes of the mass indicated, as a function of stellar mass ratio, at z=0.1. Note the feature in the merger rate on group and cluster scales.}
    \label{fig:ms_rates.pdf}
\end{figure}

To illustrate this effect, we approximate the halo merger rate shown in Fig.~\ref{fig:B/n = f(xi)} as a power law $B/n$ $\sim R_0\xi^{-1.66}M_{h,1}^{0.13}$. We then use the SHMR of \cite{Behroozi2019} to convert halo masses and mass intervals to stellar masses and mass intervals. Fig.~\ref{fig:ms_rates.pdf} shows how the merger rate is expected to vary with stellar mass ratio, for galaxies merging into systems with various primary halo masses. While the merger rate onto galaxy-mass haloes retains a simple power-law form, on group and cluster scales, the kink in the SHMR appears as a change in the slope of the merger rate in stellar mass units. This feature might be observable when recording the rate of group or galaxy-scale accretion onto clusters.

\subsubsection{Uncertainties in the Galaxy Merger Rate}

Having taken into consideration the complications discussed above in relating the galaxy merger rate to the halo merger rate, there remains the question of how accurately the latter can be determined. Galaxy mergers can be detected by either looking at objects that are very likely to merge, such as close pairs, or objects which exhibit recent evidence for merger activity, such as tidal features. 

Recent measurements of the galaxy close pair fraction in particular show that the scatter between different studies is significantly reduced if selection criteria are closely matched. \cite{Mundy.2017}, \cite{Mantha2018} and \cite{Duncan2019}, for instance, find similar trends in the merger rate as a function of redshift and stellar mass ratios $\xi$, with a scatter that is about a factor of 2--3. The combined sample also matches theoretical predictions from the Illustris hydrodynamical simulations \citep{Sublink.Rodriguez_Gomez.2015} at about this level. While this precision may improve with future work, including machine-learning (ML) approaches to identifying merging systems \citep[e.g.][]{Goulding.2018, Ackermann.2018, Bottrell.2019, Martin.2020}, the current uncertainty significantly exceeds our target accuracy of 10\%. We conclude that galaxy-scale mergers, although abundant and intrinsically interesting for the study of galaxy evolution, are unlikely to produce useful cosmological constraints.  

\section{Summary and Conclusions} 
\label{sec:p2_conc}

Tensions between current cosmological results at high and low redshift, as well as the flood of data on low-redshift clusters and galaxies expected from forthcoming surveys, encourage us to consider new methods for constraining cosmological parameters, based on non-linear structure formation and halo properties. In recent work, we found that measurements of the overall dynamical age of clusters via structural proxies such as concentration might provide quite sensitive constraints on the parameters $\Omega_m$ and $\sigma_8$. This is in part because, over a reasonable range of halo mass and redshfit, the degeneracy direction for age is almost orthogonal to the direction for abundance, and thus age constraints are very complimentary to abundance constraints.

In this paper, we have considered instead the instantaneous growth rate of haloes, as determined either from the halo merger rate, or through measures of overall accretion within the preceding dynamical time. Estimating these rates analytically, we find that halo merger rates, average growth rates, and the fraction of systems with significant recent growth (the LGS fraction) should all have slightly different dependence on the cosmological parameters, but should also be complimentary to abundance-based constraints. 

Measuring merger and growth rates in a number of different numerical simulations, we find trends similar to the analytic predictions, but do not confirm all of these exactly. Further work is needed here, to understand how mass resolution, snapshot cadence and the merger tree algorithm affect the results. Assuming the analytic predictions are correct, however, a measurement of the halo merger rate or the LGS fraction with an accuracy of $\sim$10\% would be required to distinguish between cosmologies with $\sigma_8 = 0.7$ and $\sigma_8 = 0.9$.  

Finally, we have considered several different paths to obtaining accurate measurements of the merger or growth rates observationally. On cluster scales, counting individual mergers may result in large uncertainties related to completeness, so a target of 10\% seems optimistic. On the other hand, a measurement of the LGS fraction seems more feasible, since clusters with recent episodes of significant growth should be easy to identify. Galaxy mergers provide a completely different path to determining the merger rate. There are several complications here, however, including the offset between halo merger times and galaxy merger times, and scatter in the relation between halo mass and stellar mass. Given current uncertainties in the galaxy merger rate, our target accuracy seems unrealistic on galaxy scales, although the galaxy merger rate remains extremely interesting for other reasons.

Considering these results together with those of our previous paper, we conclude that structural studies of galaxy clusters provide several promising avenues for constraining cosmological parameters. The distribution of cluster concentration parameters or projected shapes, or the prevalence of disturbed clusters showing evidence for large recent growth, should all provide tests of the cosmological model complementary to those already in use. We will continue to explore this possibility in future work.

\section*{Acknowledgements}

 JET acknowledges support from the Natural Sciences and Engineering Research Council (NSERC) of Canada, through a Discovery Grant. We thank the authors of the Illustris TNG and Bolshoi/BolshoiP simulations, and the halo finders and merger tree codes cited in section 3.1, for making their data and codes publicly available. We also thank Ravi Sheth for useful discussions about the halo merger rate, and the anonymous referee for a number of suggestions and corrections that improved the paper.

\section*{Data Availability}
$N$-body simulation data from the Bolshoi simulation is publicly available (after registration) at \url{https://www.cosmosim.org/}. $N$-body simulation data from the Illustris TNG simulation is publicly available (after registration) at \url{https://www.tng-project.org/}. The rest of the data underlying this article will be shared on reasonable request to the corresponding author.
\bibliographystyle{mnras}
\bibliography{mybib}

\appendix 
\section{Fits to the Average Halo Growth and Large Growth Fractions}
\label{sec:appendix1}
Analytical models and simulations show that the main cosmological dependence of both the average halo growth (AHG) and the fraction of haloes with large growth (LGS fraction) is through $\sigma_8$, while they depend only weakly with $\Omega_m$. In order to capture that dependence, we fit a power-law of the form 
\begin{equation}
    f(M|A, M_0, \alpha, p) = A(M/M_0)^\alpha + p
\end{equation}
for both these quantities. Note that for the average halo growth at z=1, a two-power-law function seemed more appropriate 

\begin{equation}
    f(M|A, M_0, \alpha, p) = A(M/M_0)^\alpha(1+M/M_0)^\beta+ p
\end{equation}

The parameters $A$, $M_0$, $\alpha$ and $\beta$ capture the overall shape of the function and depend on $\Omega_m$ and $z$, but are all made constant with $\sigma_8$. In contrast, the overall normalisation $p$ is the only parameter that varies with $\sigma_8$. This was motivated by the fact that in analytical models, the value of $\sigma_8$ does not affect the shape of either function. 

\begin{figure*}
    \centering
    \includegraphics[width=0.3\linewidth]{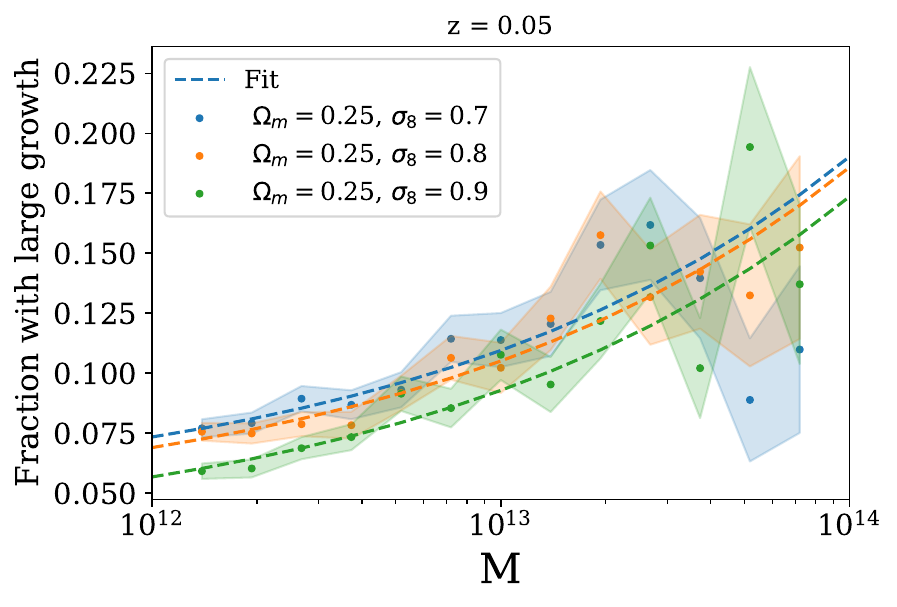}
    \includegraphics[width=0.3\linewidth]{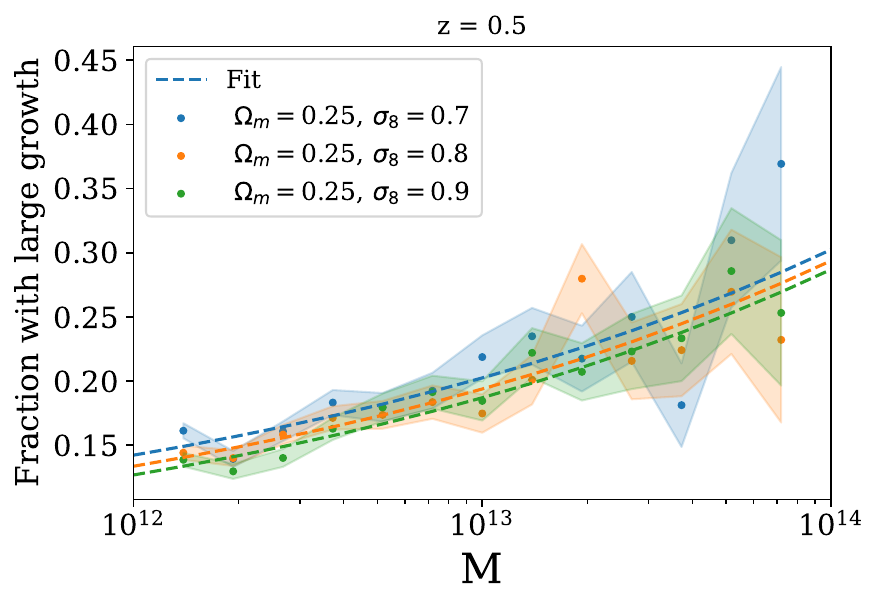}
    \includegraphics[width=0.3\linewidth]{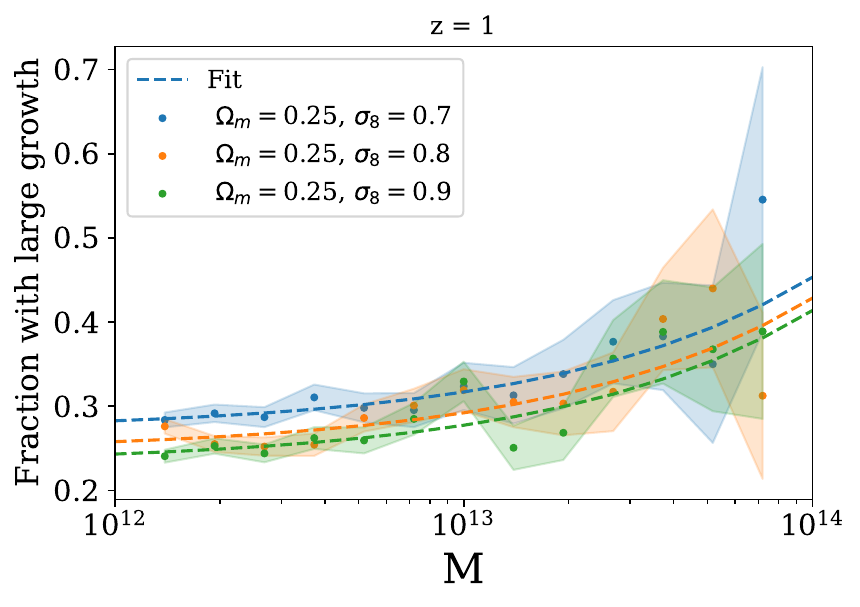} \\ 
    \includegraphics[width=0.3\linewidth]{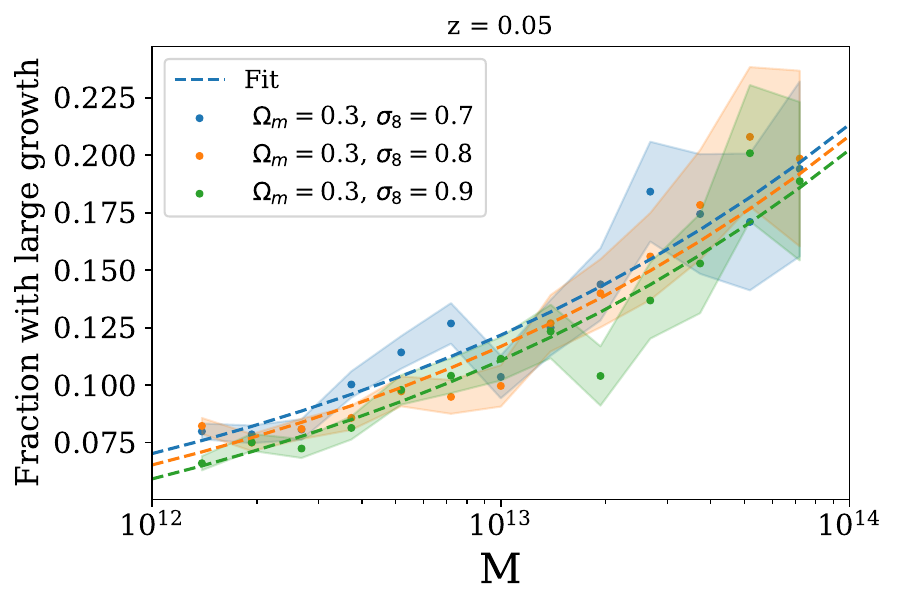}
    \includegraphics[width=0.3\linewidth]{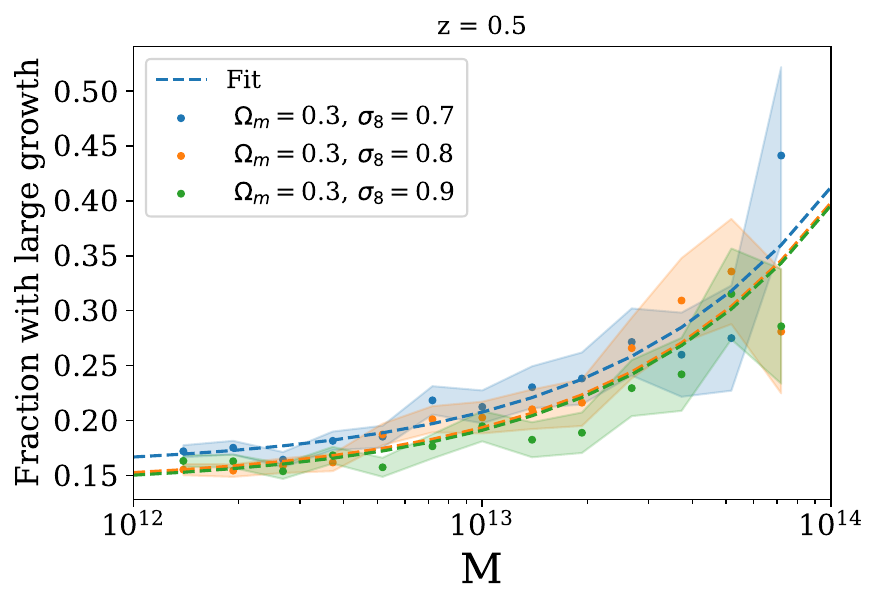}
    \includegraphics[width=0.3\linewidth]{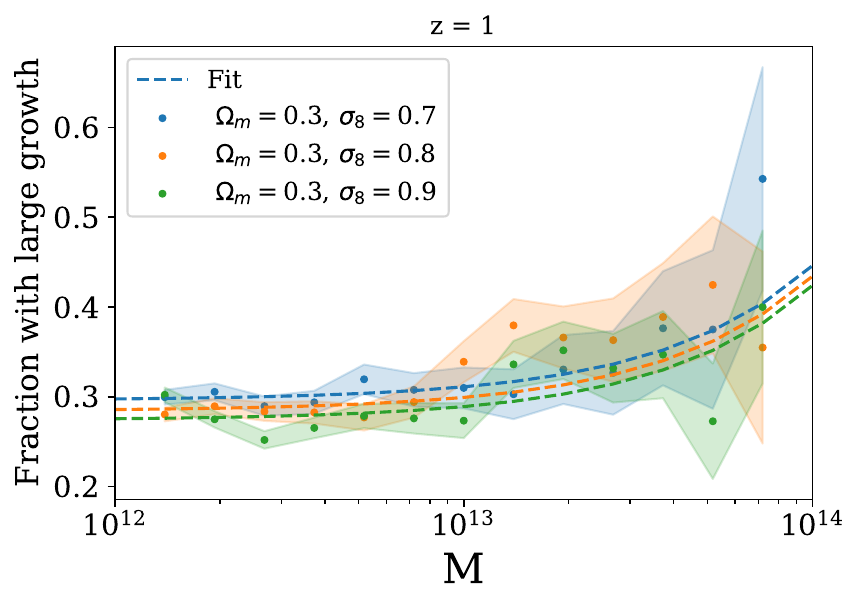} \\
    \includegraphics[width=0.3\linewidth]{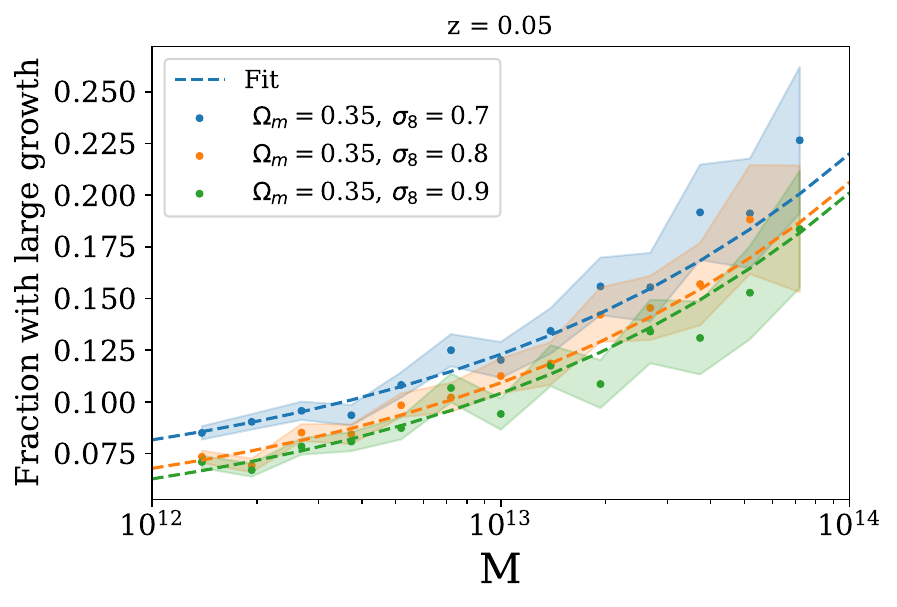}
    \includegraphics[width=0.3\linewidth]{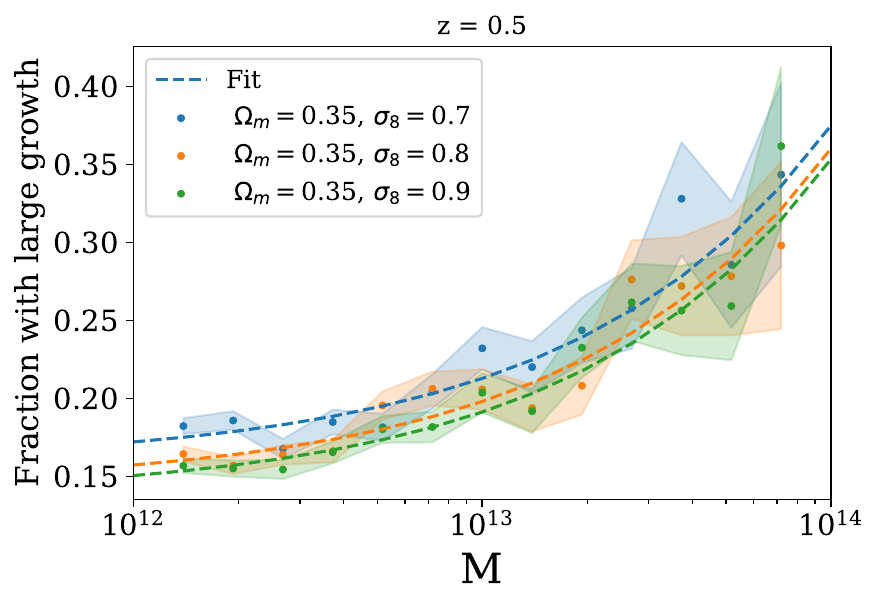}
    \includegraphics[width=0.3\linewidth]{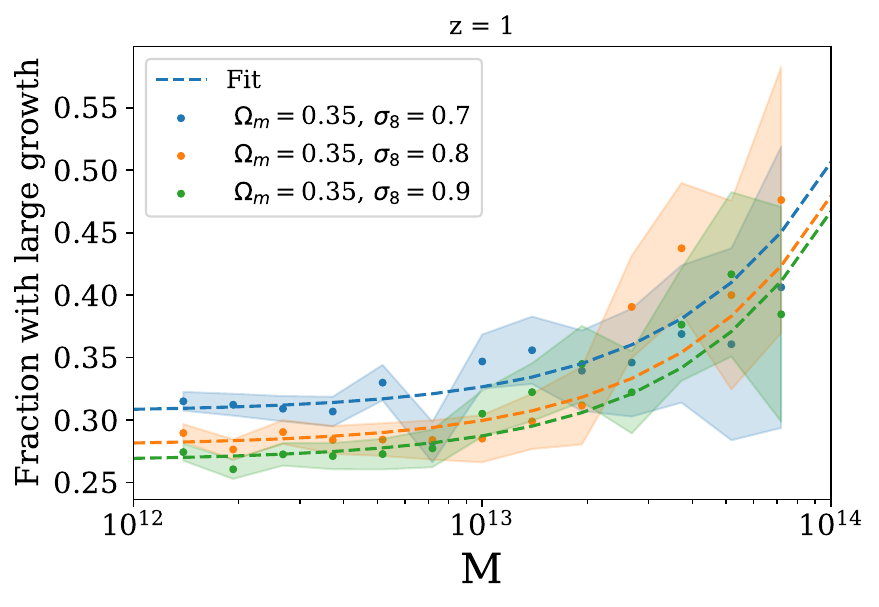}
    \caption{The cosmological dependence of the LGS fraction. Dashed lines represent power-law fits where only the normalisation between each simulation is fitted.}
    \label{fig:large_growth_sims_s8_fit}
\end{figure*}

\begin{figure*}
    \centering
    \includegraphics[width=0.3\linewidth]{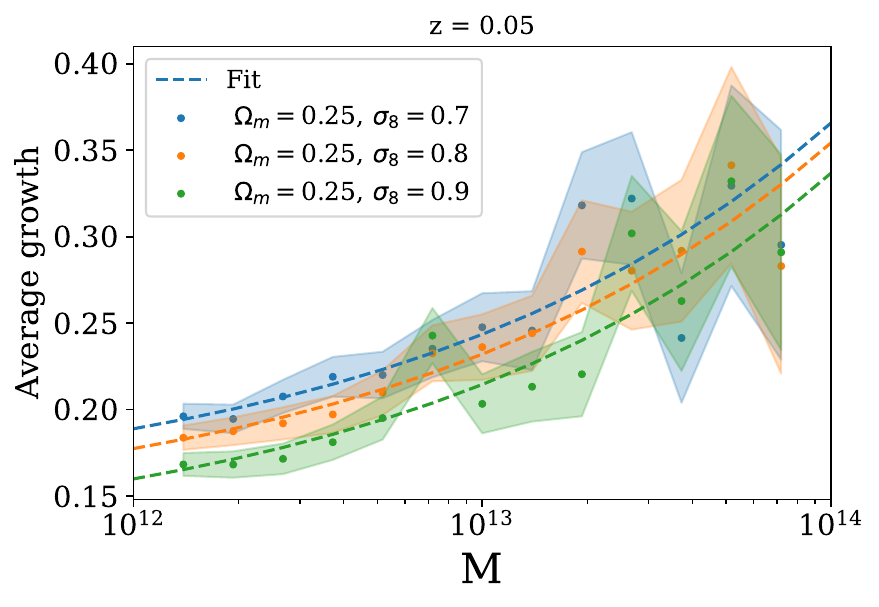}
    \includegraphics[width=0.3\linewidth]{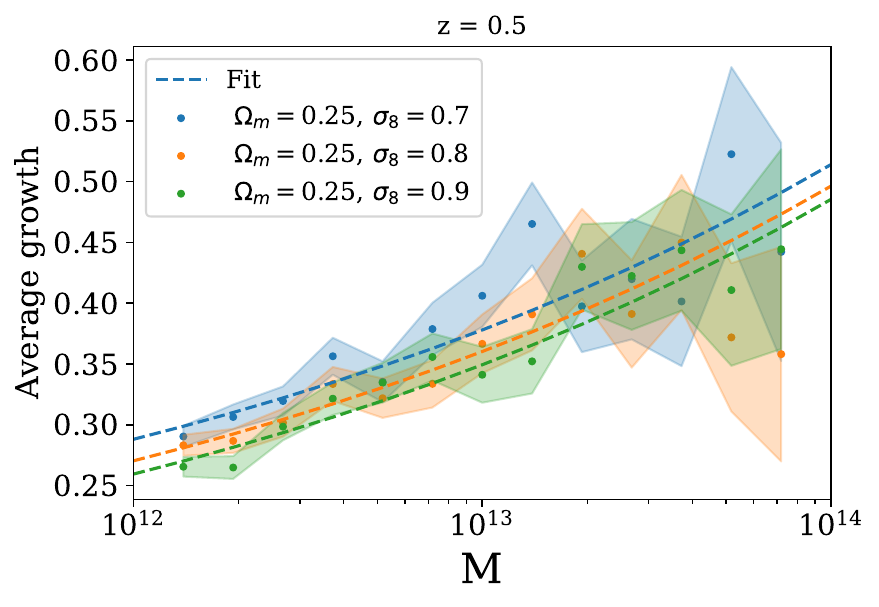}
    \includegraphics[width=0.3\linewidth]{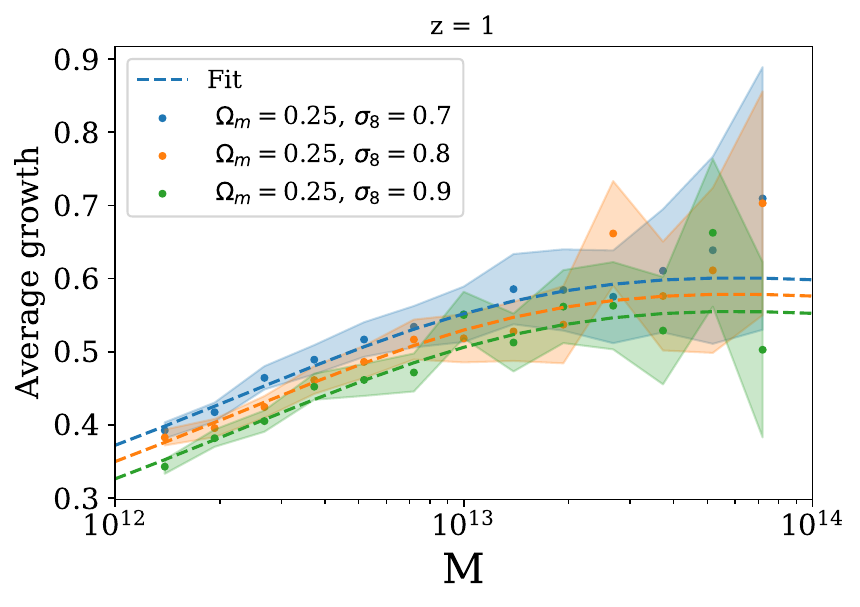} \\ 
    \includegraphics[width=0.3\linewidth]{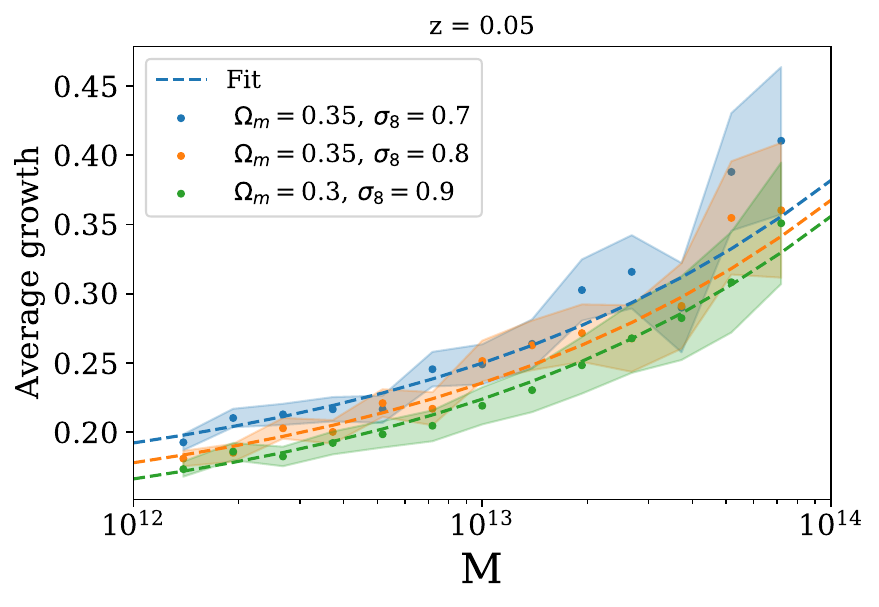}
    \includegraphics[width=0.3\linewidth]{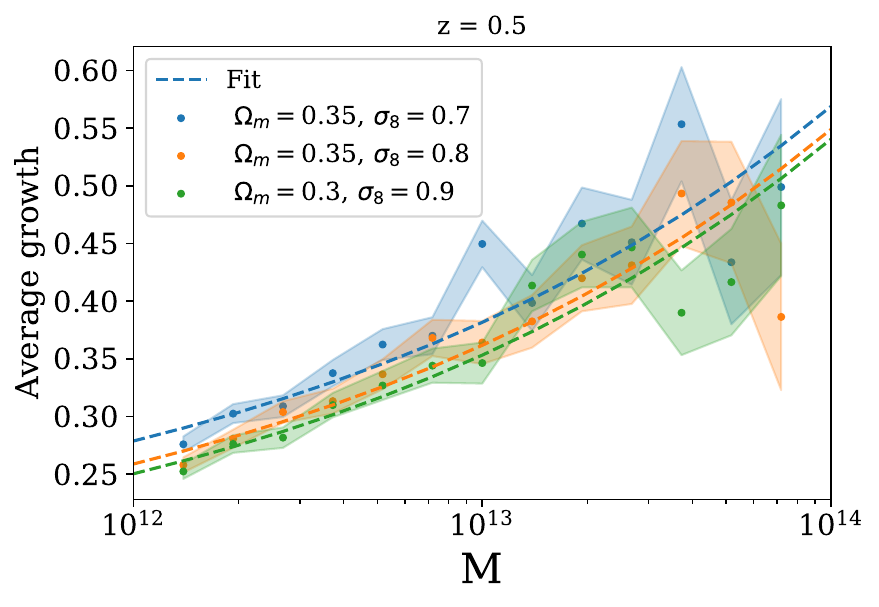}
    \includegraphics[width=0.3\linewidth]{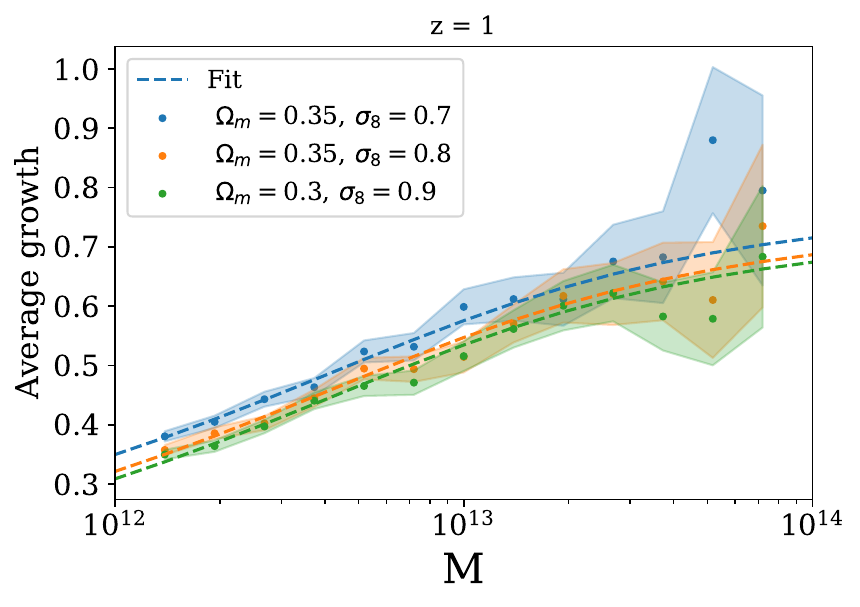} \\
    \includegraphics[width=0.3\linewidth]{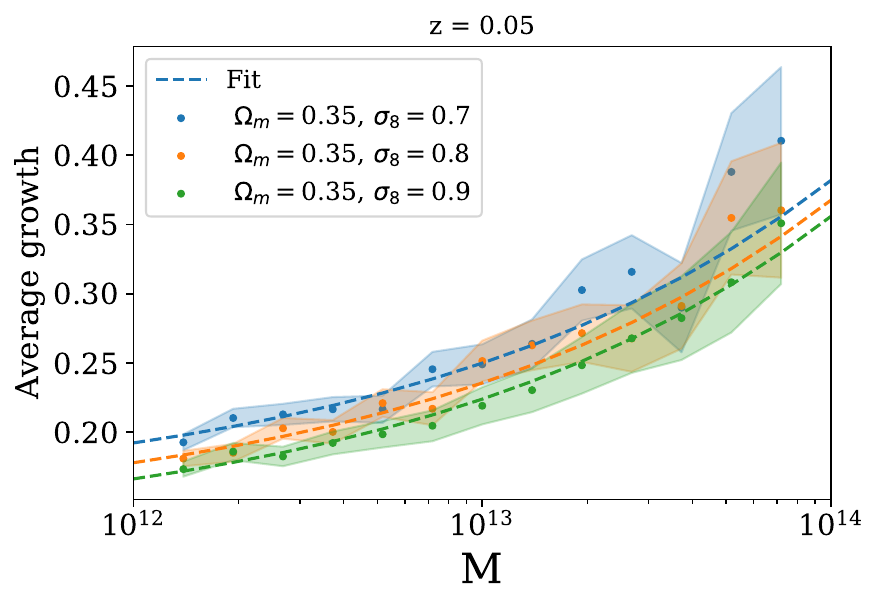}
    \includegraphics[width=0.3\linewidth]{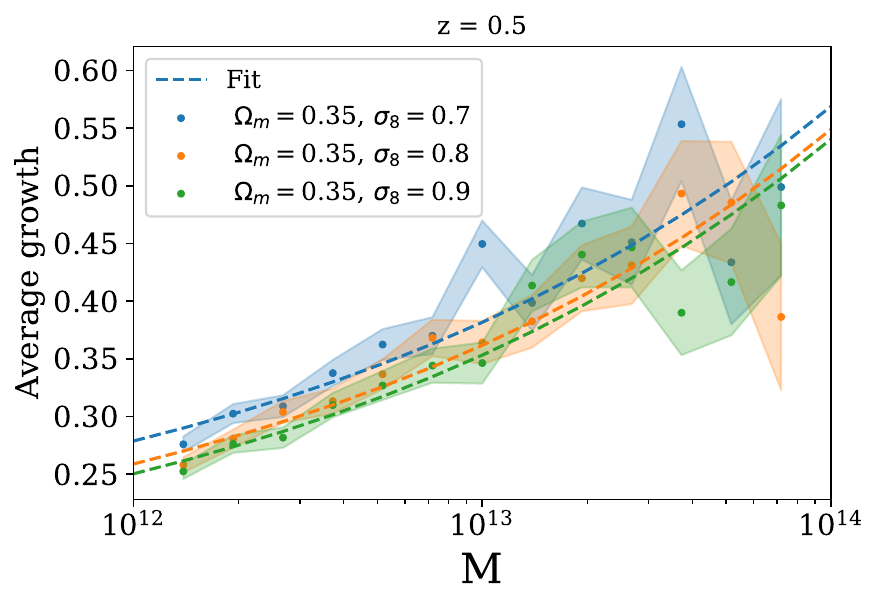}
    \includegraphics[width=0.3\linewidth]{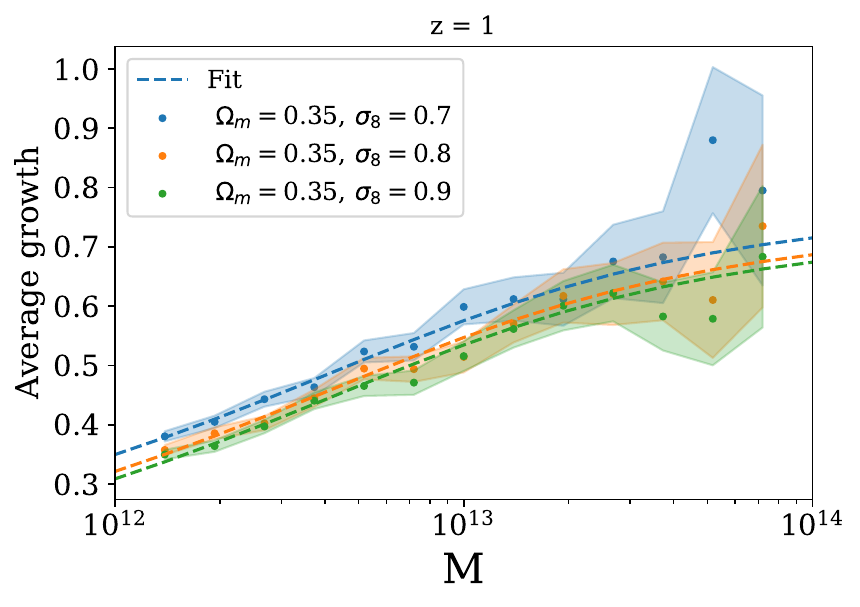}
    \caption{The cosmological dependence of the AHG. Dashed lines represent power-law fits where only the normalisation between each simulation is fitted.}
    \label{fig:av_growth_sims_s8_fit}
\end{figure*}

We show the fits to the LGS fraction and the AHG in Figs.~ ~\ref{fig:large_growth_sims_s8_fit} and \ref{fig:av_growth_sims_s8_fit} respectively, for a range of masses $10^{12}M_\odot/h<M<10^{14}M_\odot/h$.  

\bsp	
\label{lastpage}

\end{document}